\documentclass[useAMS,usenatbib]{mn2e}

\usepackage[active]{srcltx}
\usepackage{graphicx}
\usepackage{txfonts}
\usepackage{natbib}
\usepackage{multirow}
\usepackage{longtable}
\usepackage[applemac]{inputenc}
\usepackage{rotating}

\usepackage{color}
\usepackage{soul}
\definecolor{jaune}{rgb}{1.0, 1.0, 0.0}

\newcommand*{\vcenteredhbox}[1]{\begingroup
\setbox0=\hbox{#1}\parbox{\wd0}{\box0}\endgroup}

\newcommand{\kms}{km\,s$^{-1}\,$}

\newcommand{\bz}{\ensuremath{\langle B_z\rangle}}

\def\gtrsim{\mathrel{\hbox{\rlap{\hbox{\lower4pt\hbox{$\sim$}}}\hbox{$>$}}}}
\def\ltsim{\mathrel{\hbox{\rlap{\hbox{\lower4pt\hbox{$\sim$}}}\hbox{$<$}}}}

\title[NGC 1624-2: A slowly rotating, X-ray luminous Of?cp star with an extraordinarily strong magnetic field]{NGC 1624-2: A slowly rotating, X-ray luminous Of?cp star with an extraordinarily strong magnetic field\thanks{Based on spectropolarimetric observations obtained at the Canada-France-Hawaii Telescope (CFHT) which is operated by the National Research Council of Canada, the Institut National des Sciences de l'Univers (INSU) of the Centre National de la Recherche Scientifique of France, and the University of Hawaii, as well as on observations obtained using the Narval spectropolarimeter at the Observatoire du Pic du Midi (France), which is operated by the INSU. The spectroscopic data were gathered with five facilities: the 9.2 m Hobby-Eberly Telescope at McDonald Observatory (MDO), the 3.5 m Telescope at Calar Alto Observatory (CAHA), 
the 1.5 m Telescope at the Observatorio de Sierra Nevada (OSN), the 4.2 m William Herschel Telescope at the Observatorio del Roque de los Muchachos (ORM), and the 2 m Himalayan Chandra Telescope at Indian Astronomical Observatory (IAO).}}
\author[G.A. Wade et al.]{G.A. Wade\thanks{E-mail: wade-g@rmc.ca}$^1$, J. Ma\'{\i}z Apell\'aniz$^2$, F. Martins$^3$, V. Petit$^4$, J. Grunhut$^1$,  N. R. Walborn$^{5}$, 
\newauthor{R.H. Barb\'a$^{6,15}$, M. Gagn\'e$^4$, E. Garc\'{\i}a-Melendo$^{7,8}$, J. Jose$^{9}$, A.F.J. Moffat$^{10}$, Y. Naz\'e$^{11}$,}
\newauthor{C. Neiner$^{12}$, A. Pellerin$^{13}$, M. Penad\'es Ordaz$^{2}$, M. Shultz$^1$, S. Sim\'on-D\'{\i}az$^{14}$, A. Sota$^{2}$, }
\newauthor{and the MiMeS Collaboration}\\
$^{1}$Dept. of Physics, Royal Military College of Canada, PO Box 17000, Stn Forces, Kingston, Ontario K7K 7B4, Canada \\
$^{2}$Instituto de Astrof\'{\i}sica de Andaluc\'{\i}a-CSIC, Glorieta de la Astronom\'{\i}a s/n, E-18008 Granada, Spain\\
$^3$LUPM-UMR5299, CNRS \& Universit\'e Montpellier II, Place Eug\`ene Bataillon, F-34095, Montpellier, France\\
$^{4}$Department of Geology \& Astronomy, West Chester University, West Chester, Pennsylvania, USA 19383\\
$^{5}$Space Telescope Science Institute, 3700 San Martin Drive, Baltimore, MD 21218, USA\\
$^6$Instituto de Ciencias Astron\'omicas, de la Tierra y del Espacio, Casilla 467, 5400 San Juan, Argentina\\
$^7$Fundaci\'o Privada Observatori Esteve Duran, C/ Montseny 46, E-08553 Seva, Barcelona, Spain\\
$^8$Institut de Ci\`encies de l'Espai (CSIC-IEEC), Campus UAB, Facultat de Ci\`encies, Torre C5, parell, 2a pl., E-08193 Bellaterra, Spain\\
$^{9}$Indian Institute of Astrophysics, Koramangala, Bangalore, 560034, India\\
$^{10}$D\'epartement de physique, Universit\'e de Montréal, C.P. 6128, Succ. Centre-Ville, Montréal, QC, H3C 3J7\\
$^{11}$FNRS-GAPHE, D\'epartement AGO, Universit\'e de Li\`ege, All\'ee du 6 Ao\^ut 17, Bat. B5C, B4000-Li\`ege, Belgium\\
$^{12}$ LESIA, UMR 8109 du CNRS, Observatoire de Paris, UPMC, Universit\'e Paris Diderot, 5 place Jules Janssen, 92195 Meudon Cedex, France\\
$^{13}$Department of Physics and Astronomy, Texas A\&M University, College Station, TX 77843, USA\\
$^{14}$Instituto de Astrof\'{\i}sica de Canarias, E-38200 La Laguna, Tenerife, Spain\\
$^{15}$Departamento de Física, Universidad de La Serena, Chile
}

\begin{document}

\date{Accepted . Received , in original form }

\pagerange{\pageref{firstpage}--\pageref{lastpage}} \pubyear{2002}

\maketitle

\label{firstpage}

\begin{abstract}
This paper presents a first observational investigation of the faint Of?p star NGC 1624-2, yielding important new constraints on its spectral and physical characteristics, rotation, magnetic field strength, X-ray emission and magnetospheric properties. Modeling the spectrum and spectral energy distribution, we conclude that NGC 1624-2 is a main sequence star of mass $M\simeq 30~M_\odot$, and infer an effective temperature of $35\pm 2$~kK and $\log g=4.0\pm 0.2$. Based on an extensive time series of optical spectral observations we report significant variability of a large number of spectral lines, and infer a unique period of $157.99\pm 0.94$~d which we interpret as the rotational period of the star. We report the detection of a very strong - $5.35\pm 0.5$~kG - longitudinal magnetic field $\bz$, coupled with probable Zeeman splitting of Stokes $I$ profiles of metal lines confirming a surface field modulus $\langle B\rangle$ of $14\pm 1$~kG, consistent with a surface dipole of polar strength $\gtrsim 20$~kG. This is the largest magnetic field ever detected in an O-type star, and the first report of Zeeman splitting of Stokes $I$ profiles in such an object. We also report the detection of reversed Stokes $V$ profiles associated with weak, high-excitation emission lines of O~{\sc iii}, {which we propose may form in the close magnetosphere of the star.} We analyze archival Chandra ACIS-I X-ray data, inferring a very hard spectrum with an X-ray efficiency $\log L_{\rm x}/L_{\rm bol}=-6.4$, a factor of 4 larger than the canonical value for O-type stars and comparable to that of the young magnetic O-type star $\theta^1$~Ori C and other Of?p stars. Finally, we examine the probable magnetospheric properties of the star, reporting in particular very strong magnetic confinement of the stellar wind, with $\eta_*\simeq 1.5\times 10^4$, and a very large Alfven radius, $R_{\rm Alf}=11.4~R_*$.\end{abstract}

\begin{keywords}
Stars : rotation -- Stars: massive -- Instrumentation : spectropolarimetry.
\end{keywords}


\section{Introduction}

The detection \citep[e.g.][]{2009MNRAS.400L..94G}, empirical characterization \citep[e.g.][]{2011MNRAS.416.3160W} and theoretical modelling \citep[e.g.][]{2012MNRAS.tmpL.433S} of a growing sample of magnetic O-type stars is leading to a new, refined picture of the scope and impact of magnetic fields in high-mass stars. 

O-type stars are unique laboratories for investigating the physics of stellar magnetism. Magnetic fields have clear influence on their rotation rates \citep[rotation periods of most detected magnetic O stars are significantly longer than those of non-magnetic O stars of similar spectral types; e.g.][]{2009MNRAS.392.1022U, 2007MNRAS.381..433H, 2010MNRAS.407.1423M}. Evolutionary models \citep[][]{2003A&A...411..543M, 2004A&A...422..225M} and, recently, observations (Briquet et al., MNRAS, submitted) of massive stars suggest that the internal rotation profile is strongly modified by the presence of a magnetic field, enforcing essentially solid-body rotation throughout the bulk of the outer radiative zone. Magnetic fields have clear and fundamental effects on the structure, dynamics and heating of the powerful radiative winds of O stars \citep[e.g.][]{2002ApJ...576..413U, 2012MNRAS.tmpL.433S}. The lives of magnetic O-type stars are therefore expected to differ significantly from those of their non-magnetic brethren.

The subject of the present paper, NGC 1624-2\footnote{According to the numbering system of \citet{1979A&AS...38..197M}.}, is an Of?p star \citep{Walbetal10a} and the main ionizing source of the open cluster NGC 1624 (the H~{\sc ii} region S212). It is also one of only eight O-type stars in which magnetic fields have been detected with confidence. The classification Of?p was introduced by \citet{1972AJ.....77..312W} to describe spectra of early O-type stars exhibiting the presence of C~{\sc iii} $\lambda 4650$ emission with a strength comparable to the neighbouring N~{\sc iii} lines. Well-studied Of?p stars are now known to exhibit periodic spectral variations (in Balmer, He~{\sc i}, C~{\sc iii} and Si~{\sc iii} lines), narrow P Cygni or emission components in the Balmer lines and He~{\sc i} lines, and UV wind lines weaker than those of typical Of supergiants (see \citet{2010A&A...520A..59N} and references therein). With our report of a detection of a magnetic field in NGC 1624-2, magnetic fields have now been firmly detected in all of the 5 known Galactic members of this class - HD 191612: \citet{2006MNRAS.365L...6D}, \citet{2011MNRAS.416.3160W}; HD 108: \citet{2010MNRAS.407.1423M}; HD 148937: \citet{2008A&A...490..793H}, \citet{2012MNRAS.419.2459W}; CPD -28\, 2561: \citet{2012IBVS.6019....1H}, Wade et al. (2012b); NGC 1624-2: This paper - prompting the inference that there is a direct physical relationship between the magnetic field and the Of?p characteristics.

According to the recent analysis by \citet{2011MNRAS.411.2530J}, NGC 1624 is a young open cluster located significantly above the Galactic plane. Their analysis yields a heliocentric distance of $6.0\pm 0.8$~kpc \citep[in agreement with that found in the pioneering study of][]{1979A&AS...38..197M} and an age of no greater then $\sim 4$~Myr. In addition to NGC 1624-2, which is by far the brightest cluster member, 3 other apparently bright optical sources are located within 2 arcmin of the cluster centre. These sources are reported to be an early B-type main sequence star, and two probable F giants (which are probably not physically associated with the cluster).

The current paper provides a first observational characterization of NGC 1624-2, the faintest of the known Galactic Of?p stars ($V=11.8$).  In particular we report the physical parameters of the star, its rotational period, the detection of its magnetic field, and a preliminary characterization of its magnetic and magnetospheric characteristics. In Sect. 2 we describe the spectroscopic and spectropolarimetric observations upon which we base our results. In Sect. 3 we provide a short overview of the spectral properties of the star. In Sect. 4 we derive the physical properties of the star and its wind. In Sect. 5 we determine the spectral variation period of the star. In Sect. 6 we describe the magnetic field diagnosis and our constraints on the surface magnetic field. In Sect. 7 we characterize its X-ray properties based on archival data. In Sect. 8 we describe the probable magnetospheric properties of the star. In Sect. 9 we discuss our results and summarize our conclusions.

\begin{table}
\caption{\label{spectroscopy}Log of spectroscopic observations used to determine the period. $B$ refers to the spectral range that includes He\,{\sc i}~$\lambda$4471, He\,{\sc ii}~$\lambda$4542, He\,{\sc ii}~$\lambda$4686, and H$\beta$; $V$ to the range that includes 
He\,{\sc i}~$\lambda$5876; and $R$ to the range that includes H$\alpha$. J stands for the low-resolution observations from \citet{2011MNRAS.411.2530J}, G for the intermediate-resolution GOSSS observations \citep{2011hsa6.conf..467M}, and N for the high-resolution NoMaDS observations \citep{2011arXiv1109.1492M}.}
\begin{center}
\begin{tabular}{cccc | cccc}
Date       & $B$ & $V$ & $R$ & Date       & $B$ & $V$ & $R$ \\
\hline
2006-09-06 & J   & J   & J   &  2011-11-13 &     & N   & N   \\
2006-09-08 & J   & J   & J   &	 2011-11-17 &     & N   & N   \\
2007-01-26 & J   & J   & J   &	 2011-11-18 & N   &     &     \\
2008-10-14 & G   &     & G   &	 2011-11-19 & N   &     &     \\
2009-09-28 & G   &     &     &	 2012-01-15 &     & N   & N   \\
2009-09-29 & G   &     &     &	 2012-01-20 &     & N   & N   \\
2009-09-30 & G   &     &     &	 2012-01-22 &     & N   & N   \\
2009-11-01 & G   &     & G   &	 2012-01-24 &     & N   & N   \\
2009-11-03 & G   &     & G   &	 2012-01-27 &     & N   & N   \\
2009-11-17 & J   & J   & J   &	 2012-02-11 & G   & G   & G   \\
2009-11-24 & G   &     & G   &	 2012-02-15 & G   & G   & G   \\
2011-08-22 & N   & N   & N   &	 2012-03-04 &     & N   & N   \\
2011-09-09 &     & G   & G   &	 2012-03-12 & N   &     &     \\
2011-10-02 & N   & N   & N   &	 2012-03-22 &     & N   & N   \\
2011-10-03 & N   & N   & N   &	 {\bf Total N}    & 7   & 13  & 13  \\
2011-11-03 & N   &     &     &	 {\bf Total G}    & 9   &  3  &  7  \\
2011-11-10 &     & N   & N   &	 {\bf Total J}    & 4   &  4  &  4  \\
\hline
\end{tabular}
\end{center}
\end{table}

\begin{figure*}
\begin{centering}
\includegraphics[width=16cm]{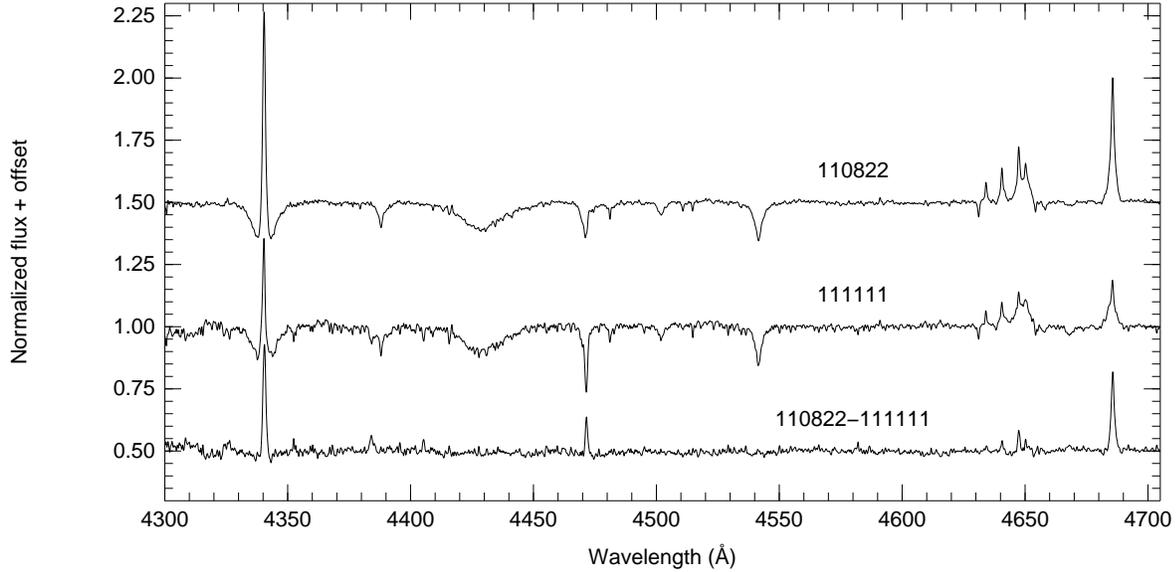}
\caption{\label{spectral lines}Two NoMaDS spectra of NGC 1624-2 (top, obtained on 22 Aug 2011, high state, phase 0.92; middle, obtained on 11 Nov 2011, low state, phase 0.43). The bottom spectrum represents the difference (high minus low). The spectra have been convolved to a resolving power $R=10\, 000$ for display purposes.}
\end{centering}
\end{figure*} 

\begin{figure}
\begin{centering}
\includegraphics[width=8cm]{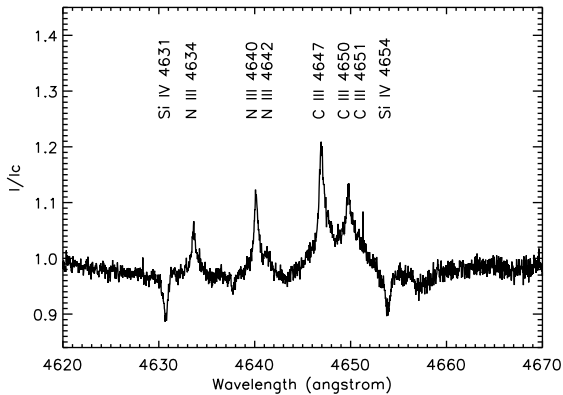}
\caption{\label{CNIII}Of?p-diagnostic emission lines of C~{\sc iii} and N~{\sc iii} near maximum emission at the full ESPaDOnS resolving power ($R=65\,000$). Note the remarkable composite emission profiles with broad and narrow components, not seen in previously investigated members of the Of?p category.}
\end{centering}
\end{figure}

\section{Observations}

\subsection{Spectroscopic observations}

The optical spectroscopic observations for the period determination were obtained within the context of three different projects. High resolution ($R=30\,000$)
spectroscopy was obtained with the NoMaDS project \citep{2011arXiv1109.1492M} using the 9.2 m Hobby-Eberly Telescope (HET). Intermediate resolution
($R=2\,500-4\,000$) spectroscopy was obtained within the GOSSS project \citep{2011arXiv1109.1492M} using the 1.5 m telescope of the Observatorio de Sierra Nevada (OSN), the 3.5 m telescope of the Calar Alto Observatory (CAHA), and the 4.2 m William Herschel Telescope (WHT). Finally, low resolution ($R=1\,500$) spectroscopy was obtained by \citet{2011MNRAS.411.2530J} using the 2 m Himalayan Chandra Telescope. 

The 33 NoMaDS spectra were gathered using the high resolution fiber-fed echelle spectrograph (HRS) at the HET. The spectra were obtained with two different instrument setups (600g4739K and 600g6302) allowing us to cover the entire wavelength range of $\sim$3800-7300~\AA. The fiber sdiamter was 2$\arcsec$. We obtained two exposures of 3600 seconds in the blue-setup ($\sim$3800-5700~\AA) and two of 1800 seconds for the red-setup ($\sim$5400-7300~\AA). The data were reduced following the IRAF/package-based procedure suggested on the HET website\footnote{http://hydra.as.utexas.edu/?a=help\&h=29, ver. 22 Feb 2012}. Cosmic rays were removed using the L.A. Cosmic package \citep{2001PASP..113.1420V}. The spectra were flux normalized as well as wavelength calibrated. The typical resulting signal-to-noise ratio (SNR, per resolution element) is between 150 and 250, depending on the wavelength range. For NGC\,1624-2, additional spectra were obtained in order to remove the nebular contribution. For these observations the fiber was positioned about 14$\arcsec$ NE and SW of NGC\,1624-2 to avoid stellar contamination, and the spectra were reduced as described above and subtracted from the stellar observations.

The 19 GOSSS spectra were obtained with various spectrographs fed by the OSN, CAHA and WHT telescopes as part of the larger Galactic O-Star Spectroscopic Survey. The data {were} reduced using a pipeline specifically written for the project, described by \citet{Sotaetal11a}.

A log of the 64 spectroscopic observations is presented in Table~\ref{spectroscopy}, where we specify which regions were observed on each date.

\subsection{Spectropolarimetric observations}

Five high resolving power ($R\simeq 65\,000$) spectropolarimetric (Stokes $I$ and $V$) observations of NGC 1624-2 were collected with ESPaDOnS at the Canada-France-Hawaii Telescope (CFHT) between Feb 1 and 9 2012. An additional observation was obtained with Narval (essentially a twin of ESPaDOnS) on the Bernard Lyot telescope (TBL) on Mar 24 2012. All of the spectropolarimetric observations were obtained within the context of the Magnetism in Massive Stars (MiMeS) Large Programs. Each spectropolarimetric sequence consisted of four individual subexposures, each of 600 s duration (for ESPaDOnS) or 1\,200 s duration (for Narval), taken in different configurations of the polarimeter retarders. From each set of four subexposures we derived Stokes $I$ and Stokes $V$ spectra following the double-ratio procedure described by \citet{1997MNRAS.291..658D}, ensuring in particular that all spurious signatures were removed to first order. Null polarization spectra (labeled $N$) were calculated by combining the four subexposures in such a way that polarization cancels out, allowing us to verify that no spurious signals are present in the data \citep[see][for more details on the definition of $N$]{1997MNRAS.291..658D}. All frames were processed using the automated reduction package Libre ESpRIT \citep{1997MNRAS.291..658D}. The peak SNRs per 2.6 \kms\  velocity bin in the reduced spectra range from 86-140, with the variation due principally to seeing conditions. 


\begin{figure*}
\centering
\vcenteredhbox{\includegraphics[width=9.5cm]{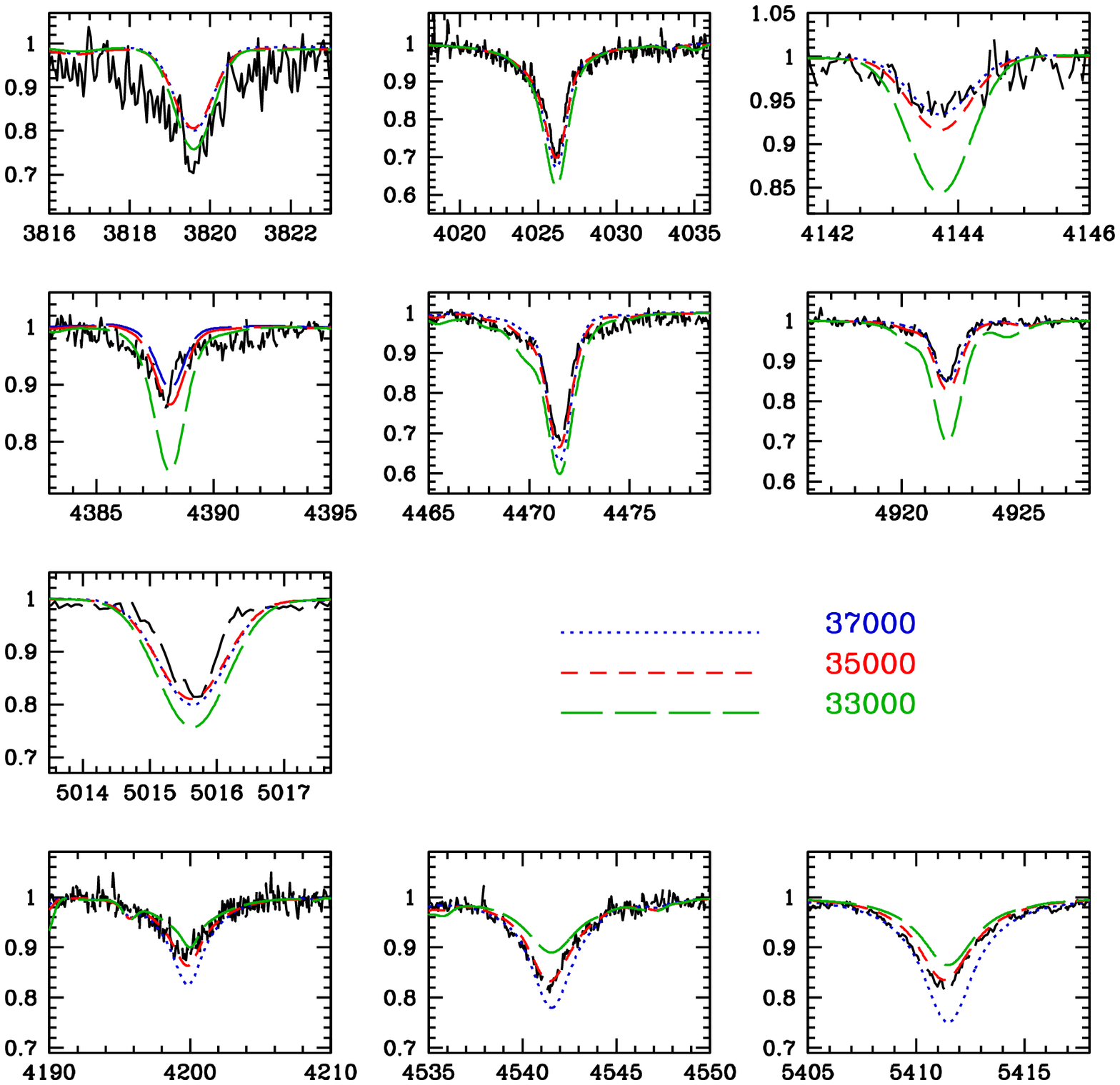}}\vcenteredhbox{\includegraphics[width=8cm]{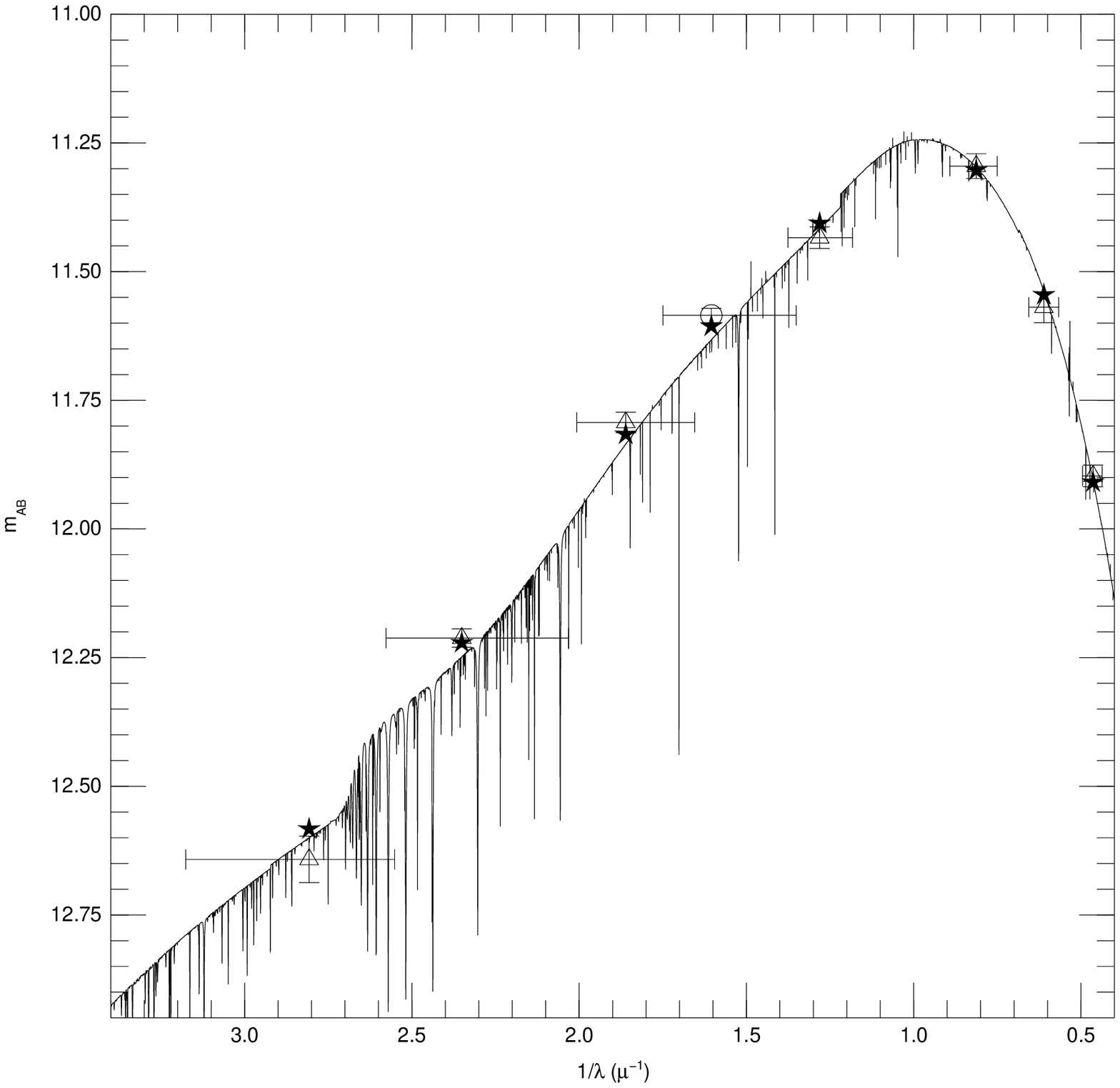}}
\caption{\label{paramsfig}{\em Left panel: }\ {Fit of the observed He lines (black solid line) by three CMFGEN models with $T_{\rm eff}=33\,000$ (green long-dashed line), 35\,000 (red short-dashed line) and 37\,000 K (blue dotted line).  {\em Right panel:}\ Best SED (black solid curve) from the CHORIZOS run with fixed $T_{\rm eff}=35\,000$~K. The {triangles with error bars} are the photometric magnitudes used in the fit ($UBVIJHK$), {the circle represents the measurement} that was unused ($R$), and {the stars} are the synthetic magnitudes corresponding to the best SED.}}
\end{figure*}

\section{The optical spectrum of NGC 1624-2}
\label{spectrum}

Fig.~\ref{spectral lines} shows the blue-green spectra of NGC~1624-2 from the HET/NoMaDS data (convolved with a Gaussian to a uniform $R = 10\,000$ for classification purposes) near maximum and minimum emission phases, together with their difference.  The apparent spectral types from the He ionization ratios are O6.5 and O8, respectively, although both are affected by particular spectral features: at maximum, longward emission filling in He~{\sc i} $\lambda4471$ is clearly seen, while at minimum, the He~{\sc i} and He~{\sc ii} profiles are very different due to the stronger Stark wings in the latter, and indeed, the equivalent width ratio $\sim1$ corresponds to a type near O7.    

Fig.~\ref{CNIII} displays the N~{\sc iii} $\lambda4640$, C~{\sc iii} $\lambda4650$ complex near maximum emission at the full ESPaDOnS resolving power of 65\,000. NGC 1624-2 exhibits the most extreme Of?p spectrum known to date, with the C~{\sc iii} emission lines {\it stronger} than those of N~{\sc iii}, and stellar Balmer emission much stronger than in any other. The N~{\sc iii} and C~{\sc iii} features (as well as, we suspect, He~{\sc ii} $\lambda4686$) display remarkable composite emission profiles with broad and narrow components, not seen in previously investigated members of the Of?p category.  We hypothesize that the broad components are photospheric as in normal Of or Ofc stars \citep{Walbetal10a}, while the narrow components typical of Of?p spectra are magnetospheric.  In agreement with this interpretation, it is seen in Fig.~\ref{spectral lines} that only the narrow components vary with phase.  However, the narrow C~{\sc iii} emission does not disappear at minimum, so unlike most other members of the category NGC~1624-2 remains Of?p at both extreme phases. Moreover, the apparently greater intensity of C~{\sc iii} as compared to N~{\sc iii} is largely due to the broad component, so that the intensities of both species in the narrow system are more comparable, similarly to other Of?p spectra.

This is the first case combining Ofc and Of?p emission profiles, for which we introduce the new notation Of?cp;  thus the full classifications of these spectra are now O6.5f?cp and O8f?cp at maximum and minimum, respectively.  As discussed by \cite{Walbetal10a}, the Ofc phenomenon in normal Galactic spectra is strongly peaked toward type O5.  However, the optical component of the high-mass X-ray binary HD~153919 (2U 1700-37) is O6.5~Iafc \citep[see, e.g., ][A. Sota et al. in prep.]{1975ApJ...200..133C}, as is the primary of the LMC double-lined spectroscopic binary R139 \citep{2011A&A...530L..10T}.  Thus, the Ofc phenomenon also occurs at later spectral types in some anomalous objects. 


\section{Stellar physical and wind properties}
\label{physicalproperties}

The atmospheric parameters of NGC1624-2 have been derived by means of spectroscopic analysis with the code CMFGEN, fitting the minimum-emission state spectrum (the average of NoMaDS spectra acquired on 17/18 Nov 2011). A full description of CMFGEN is provided by \citet{hm98}. In brief, it computes non-local thermodynamic equilibrium (non-LTE) atmosphere models including winds and line--blanketing. The statistical equilibrium and radiative transfer equations are solved in the comoving frame through an iterative scheme to compute the level populations and radiation field. The temperature structure is set by the condition of radiative equilibrium. Our models include the following elements; H, He, C, N, O, Ne, Si, S, Ar, Ca, Fe, Ni. The solar abundances of \citet{ga07} are used unless stated differently. The velocity structure is specified at the beginning of the computation. It is usually a combination of a pre-computed photospheric structure connected smoothly to a so--called $\beta$ velocity law ($v \propto (1-r/R_{*})^{\beta}$ where $r$ is the radial coordinate and $R_{*}$ the stellar radius). This structure is subsequently iterated to satisfy the momentum equilibrium equation in the inner atmosphere. Once the atmosphere model is converged, a formal solution of the radiative transfer equation is performed to yield the emergent spectrum. A microturbulent velocity varying with height from 10 to 200 km s$^{-1}$ {(beginning in the photosphere and continuing to 100~$R_\odot$)} is used. 

Prior to a comparison between synthetic and observed spectra, it is necessary to determine the projected rotational velocity and macroturbulent broadening. We used the Fourier transform method for the former \citep[e.g.][]{1995POBeo..50...75J, sergio06}, focusing on the C~{\sc iv} $\lambda$5801 line. The Fourier transform did not show any zero before reaching the noise level, implying an upper limit on $v\sin i$ of $\sim 15$~km s$^{-1}$. Adopting this value, we convolved our synthetic spectra with a rotational profile and a Gaussian profile aimed at reproducing isotropic macroturbulence. We found that a macroturbulent velocity of about 25 km s$^{-1}$ gave the best fit to the observed He lines. We note based on the results of Sect.~\ref{magnetic} that this additional broadening is likely to be magnetic in origin, and so isotropic macroturbulence will provide only an approximate reproduction of its effects. The combined line broadening (rotational, turbulent, magnetic) of NGC 1624-2 is comparable to that of the very sharp-lined magnetic O9IV star HD 57682 \citep[][and Grunhut et al., MNRAS, submitted]{2009MNRAS.400L..94G} . 

The He~{\sc i}/He~{\sc ii} ionization balance was used to derive $T_{\rm eff}$. As for the other Of?p stars, we relied on the spectrum closest to the minimum-emission state. It has the least contaminated He~{\sc i} lines. Based on the absence of significant variability of the majority of the He~{\sc ii} lines in our spectra, we assume these lines are principally photospheric. Fig.~\ref{paramsfig} (left panel) shows the He lines (7 lines of He~{\sc i}, 3 lines of He~{\sc ii}; He~{\sc ii} $\lambda 4686$ was not used, as it is strongly variable) used in our analysis together with three selected models at 33 kK, 35 kK and 37 kK. Overall, the 35~kK model provides the best representation of the observed profile. The He~{\sc ii} lines are either too weak or too strong if $T_{\rm eff}=37$~kK or 33~kK respectively. The He~{\sc i} lines also appear too strong at low $T_{\rm eff}$. However, we note that the exact shape of the He~{\sc i} lines is uncertain (due to residual contamination by unmodelled contributions from the non-spherical component of the wind) and consequently, a safe uncertainty on $T_{\rm eff}$ is 2000 to 3000 K.

The surface gravity of NGC 1624-2 was derived from the shape of the Balmer-line wings, as is usually done for O stars. We found that a value of $\log g=4.0$ was the best compromise to reproduce all lines. A slightly higher value ($\log g=4.2$) is indicated by H$\beta$ and H$\epsilon$ compared to H$\gamma$. We can safely exclude any value below 3.7, the broadening being insufficient to reproduce the observed profiles in that case. We thus adopt a value of $\log g=4.0\pm 0.2$.  


\begin{table}
\centering
\caption{\label{params}Results of the CHORIZOS modelling of the spectral energy distribution. The first column gives the results for the run in which the luminosity class (LC) was allowed to vary, but in which the effective temperature ($T_{\rm eff}$) was fixed to 4.8, i.e. close to luminosity class 5. The second column gives the results for the run in which the effective temperature was allowed to vary, but in which the luminosity class (LC) was fixed.}
\begin{tabular}{lcc}
\multicolumn{1}{c}{Quantity} & Free LC               & Free $T_{\rm eff}$ \\
\hline
$\chi^2_{\rm red}$           & 2.03                  & 0.86               \\
$T_{\rm eff}$ (K)            & 35\,000               & 31\,000 $\pm$ 2800 \\
$\log g$ (cgs)               & 3.891 $\pm$ 0.140     & 3.957 $\pm$ 0.022  \\
$\log L$ (solar)             & 5.125 $\pm$ 0.171     & 4.632 $\pm$ 0.214  \\
$E(4405-5495)$ (mag)         & 0.802 $\pm$ 0.015     & 0.794 $\pm$ 0.018  \\
$E(B-V)$ (mag)               & 0.802 $\pm$ 0.014     & 0.792 $\pm$ 0.016  \\
$R_{5495}$                   & 3.752 $\pm$ 0.092     & 3.736 $\pm$ 0.096  \\
$A_V$ (mag)                  & 3.040 $\pm$ 0.031     & 2.991 $\pm$ 0.046  \\
$\log d$ (pc)                & 3.712 $\pm$ 0.094     & 3.537 $\pm$ 0.059  \\
\hline
\end{tabular}\end{table}

\begin{table}
\centering
\caption{\label{param_summary}Summary of adopted stellar, wind, magnetic and magnetospheric properties of NGC 1624-2. The effective temperature $T_{\rm eff}$, surface gravity $\log g$, radius $R_\star$, luminosity $L_\star$ and mass $M_\star$, as well as the reddening $E(B-V)$, extinction $R_{\rm V}$ and distance $\log d$ were determined based on modelling of observations described in Sect.~\ref{physicalproperties}. The upper limit on the projected rotational velocity - determined from the inferred radius and period assuming rigid rotation - is consistent with the negligible additional line broadening required to fit the magnetically-split profiles of the C~{\sc iv} line profiles (Sect.~\ref{magnetic}). The wind mass-loss rate $\dot M$ and terminal velocity $v_\infty$ are calculated based on theoretical considerations in Sect.~\ref{magnetosphere}. The magnetic field dipole surface strength $B_{\rm d}$ was estimated based on measurement and modelling of Zeeman effect, described in Sect.~\ref{magnetic}. Finally, the wind magnetic confinement parameter $\eta_\star$, the Alfven radius $R_{\rm Alf}$, the rotation parameter $W$ and the Kepler radius $R_{\rm Kep}$ are calculated based on measured and inferred quantities in Sect.~\ref{magnetosphere}.}
\begin{tabular}{l|ll}
\hline
Spectral type &  O6.5f?cp-O8f?cp            \\
$T_{\rm eff}$ (K) & 35 000 $\pm$ 2000 \\
log $g$ (cgs) & 4.0 $\pm$ 0.2     \\
R$_{\star}$ (R$_\odot$) & $10\pm 3$ \\
$v\sin i$ (km\,s$^{-1}$) & $\ltsim 3$  \\
$\log (L_\star/L_\odot)$ & $5.10\pm 0.2$   \\
$M_{\star}^{\rm spec}$ ($M_{\odot}$) & $34\pm 31$ \\
{$M_{\star}^{\rm evol}$ ($M_{\odot}$)} & {$28^{+7}_{-5}$} \\
\hline
{$\log \dot{M}$} (M$_{\odot}$\,yr$^{-1}$) & $-6.8$  \\
$v_{\infty}$ (km\,s$^{-1}$) & 2875  \\
\hline
$E(B-V)$ (mag) &$0.802\pm 0.02$ \\
$R_{\rm V}$ & $3.74\pm 0.1$ \\
$\log d$ (pc)  & 3.712 $\pm$ 0.1 \\
\hline
$B_{\rm d}$ (kG) & $\sim 20$ \\
$\eta_\star$ & $1.5\times 10^4$ \\
R$_{\rm Alf}$ ($R_*$) & $11.4$ \\
$W$ & $4\times 10^{-3}$ \\
R$_{\rm Kep}$ ($R_*$) & $>40$ \\
$\tau_{\rm spin}$ (Myr) & 0.24\\
\hline\hline
\end{tabular}
\end{table}

We also investigated the nitrogen content of NGC 1624-2 by means of its N~{\sc iii} lines \citep[e.g.][]{2012A&A...538A..29M}. We used the N~{\sc iii} $\lambda$4195, N~{\sc iii} $\lambda$4510--4525 and N~{\sc iii} $\lambda$4535 features, as all other nitrogen lines are either in emission, absent, or of very low SNR (below 10). Assuming the adopted microturbulence described above, a value of N/H=$(1.0\pm 0.4)\times 10^{-4}$ ({by number}) gives a satisfactory fit. \citet{2012A&A...538A..29M} tested the effect of a change of the microturbulent velocity in the atmosphere model computation (as well as the output synthetic spectrum) and found that a 5~\kms increase in the microturbulence resulted in a 10-20\% change in the derived abundance. This effect, while not negligible, is smaller than the effect of $T_{\rm eff}$ uncertainties. Ultimately, while the exact value of the N abundance is uncertain due to the turbulent/magnetic and temperature uncertainties, we can conclude that the star is not strongly N enriched {([N/H]$\ltsim$0.3)}. 

In the absence of UV spectroscopy, and given the peculiar shape of the emission in wind sensitive optical lines (He~{\sc ii} $\lambda$4686, H$\alpha$), it is not possible to derive the mass loss rate of NGC 1624-2 based on the observations. First, the terminal velocity is not known (it is usually derived from the blueward extension of UV P-Cygni profiles). Secondly, the optical emission lines are very narrow, as in the other Of?p lines, and are probably the result of a complex (magnetically-confined) wind geometry that cannot be reproduced by our 1D models \citep{2012MNRAS.tmpL.433S}. 

In order to derive the distance and extinction of NGC 1624-2 we used the Bayesian (spectro)photometric code CHORIZOS 
\citep{Maiz04c}. In the latest CHORIZOS version, the user can select distance to be an independent 
parameter by applying atmosphere models \citep[TLUSTY for OB stars;][]{LanzHube03,LanzHube07} calibrated in luminosity with
the help of Geneva stellar evolutionary tracks (excluding rotational effects). In such models the intrinsic parameters are effective temperature 
($T_{\rm eff}$) and photometric luminosity class (LC, defined in a similar way to its spectroscopic counterpart, with 
0.0 corresponding to hypergiants and 5.5 to ZAMS). The extrinsic parameters are reddening ($E(4405-5495)$, the 
monochromatic equivalent of $E(B-V)$), extinction law ($R_{5495}$, the monochromatic equivalent of $R_V$), and 
logarithmic distance ($\log$ d). CHORIZOS uses the family of extinction laws of Ma\'{\i}z Apell\'aniz et al. (2012, in 
preparation), which improve upon the \citet{Cardetal89} laws by using spline interpolation in the optical range 
instead of a seventh-degree polynomial and by correcting the extinction in the $U$ band. 

As input photometry, we used the $UBVI$ magnitudes of \citet{2011MNRAS.411.2530J} and the $JHK$ magnitudes of 2MASS \citep{Skruetal06}. 
We discarded the $R$ magnitude of \citet{2011MNRAS.411.2530J} due to the influence of the 
H$\alpha$ emission on the stellar SED, though we later checked that its inclusion would not have greatly changed our results. We did two 
different runs: in the first one, we fixed $T_{\rm eff}$ to be 35\,000 K (the value derived from spectroscopy) and 
allowed LC to vary in such a way that $3.7 < \log g < 4.3$ ({slightly larger than the range derived from spectroscopy}). In the second run, we
left $T_{\rm eff}$ as a free parameter and fixed LC to be 4.808 (i.e. close to the spectroscopic luminosity class V), 
since that value corresponds exactly to $\log g$ = 4.00 (cgs) for $T_{\rm eff}$ = 35\,000 K and $\log g$ is 
approximately constant for a fixed LC for O stars near LC = 5.0. In the two runs we left the three extrinsic parameters 
(reddening, extinction law, and logarithmic distances) free. 

The results of the two CHORIZOS runs are shown in Table~{\ref{params}} and the mean SED from the first run is shown in Fig.~{\ref{paramsfig}}. The 
two runs show good values of the reduced $\chi^2$, indicating that the photometry is consistent with the input SED 
models and extinction laws. All of the results are consistent (within 2$\sigma$) between the two runs, with the
values for $E(4405-5495)$, $E(B-V)$, $R_{5495}$, and $A_{\rm V}$ practically indistinguishable. The most significant differences 
between the two runs are in $T_{\rm eff}$, $\log L$, and $\log d$: the first run indicates a hotter, less luminous, and
closer object in comparison with the second run. The second run has a better 
$\chi^2_{\rm red}$ but that can be ascribed to a random fluctuation in the input photometry (the obtained 
$T_{\rm eff}$ is within 2$\sigma$ of the spectroscopic value) or a small effect of the circumstellar material in the 
SED\footnote{We did equivalent CHORIZOS 
runs with the \citet{Cardetal89} extinction laws and we found the same effect as in our derivation of the new
Ma\'{\i}z Apell\'aniz extinction laws: for the fixed $T_{\rm eff}$ case $\chi^2_{\rm red}$ is larger and for the free 
$T_{\rm eff}$ case the measured value is further offset from the spectroscopic one.}. The distance obtained in the first run ($5.2 \pm 1.1$ kpc) is in better agreement with the values from the 
literature ($\sim 6$~kpc). The values for the extinction indicate that $A_{\rm V}$ is close to 3 magnitudes and that the value of $R_{5495}$ 
is slightly larger than the mean Galactic value of 3.1-3.2, something that is typical for objects in H\,{\sc ii} regions. 

Using the CHORIZOS-derived distance, reddening and extinction, the best CMFGEN fit to the SED corresponds to $\log L/L_\odot=5.08$. With the quoted uncertainty on $d$, the uncertainty on $\log L/L_\odot$ is about 0.15 dex.  Therefore the CMFGEN ($5.08\pm 0.15$) and CHORIZOS ($5.13\pm 0.17$) values of $\log L/L_\odot$ are in excellent agreement. 

Based on the derived temperature ($T_{\rm eff}=35\, 000\pm 2000$~K and luminosity (averaging the CMFGEN and CHORIZOS results, $\log L_*/L_\odot=5.1\pm 0.2$), we estimate the stellar radius $R_*=10\pm 3~R_\odot$. {The mass of NGC 1624-2 derived from the spectroscopic analysis (i.e. from $\log g$  and the derived radius) is $M_*=34\pm 31~M_\odot$. The uncertainty is very large because the error on $\log g$ is nearly a factor of 1.6 (0.2 dex). This "spectroscopic" mass can be compared to the "evolutionary" mass derived simply from the ($T_{\rm eff}, L_*$) position of NGC 1624-2 on the HR diagram. Using the evolutionary tracks of \citet{2005A&A...429..581M}, we find $M_{\rm evol}=28^{+7}_{-5}~M_{\odot}$. This estimate implicitly assumes that the evolutionary tracks used can represent the evolution of the star. Given the peculiar properties of NGC 1624-2 (in particular its very strong magnetic field, very slow rotation and probably modified mass loss) this is not completely obvious, and thus the evolutionary mass should be regarded as only indicative.}

The adopted parameters for NGC 1624-2 are summarized in Table~\ref{param_summary}.

\begin{figure}
\begin{centering}
\includegraphics[width=16cm]{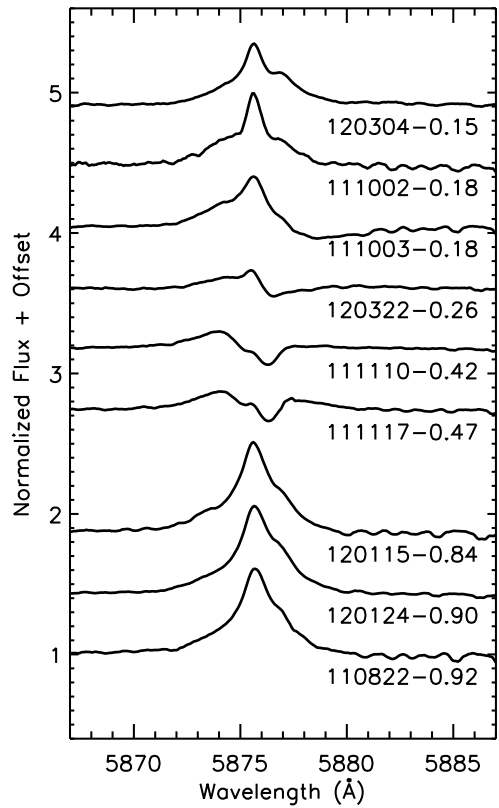}
\caption{\label{5876}Illustration of the variation of the He~{\sc i} $\lambda 5876$ line in the NoMaDS spectra. Numbers associated with each spectrum indicates date in the format yymmdd, and the phase computed according to Eq. (1).}
\end{centering}
\end{figure}

\begin{figure*}
\centering
\includegraphics[width=7.7cm]{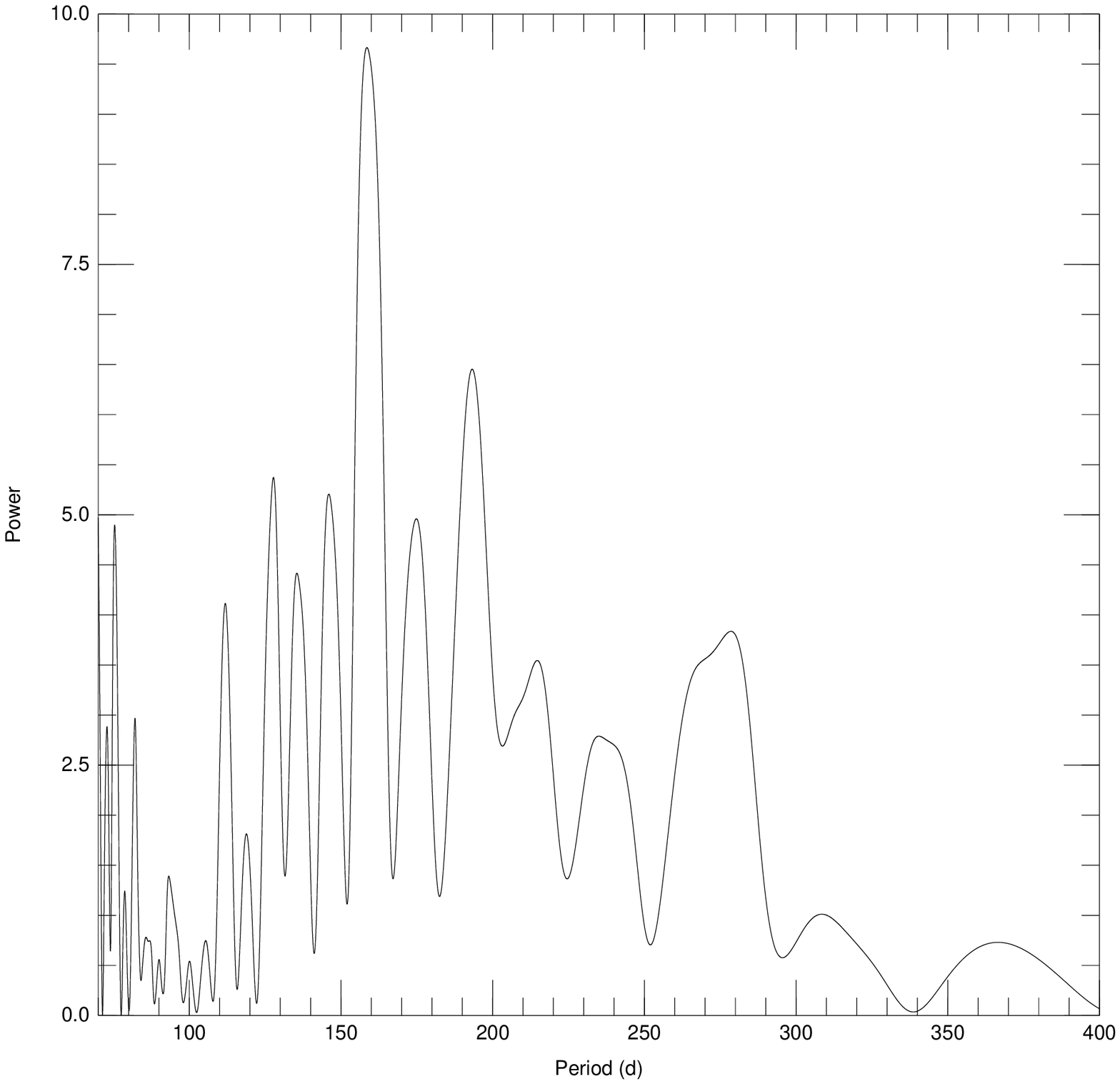}\hspace{0.5cm}\includegraphics[width=7.7cm]{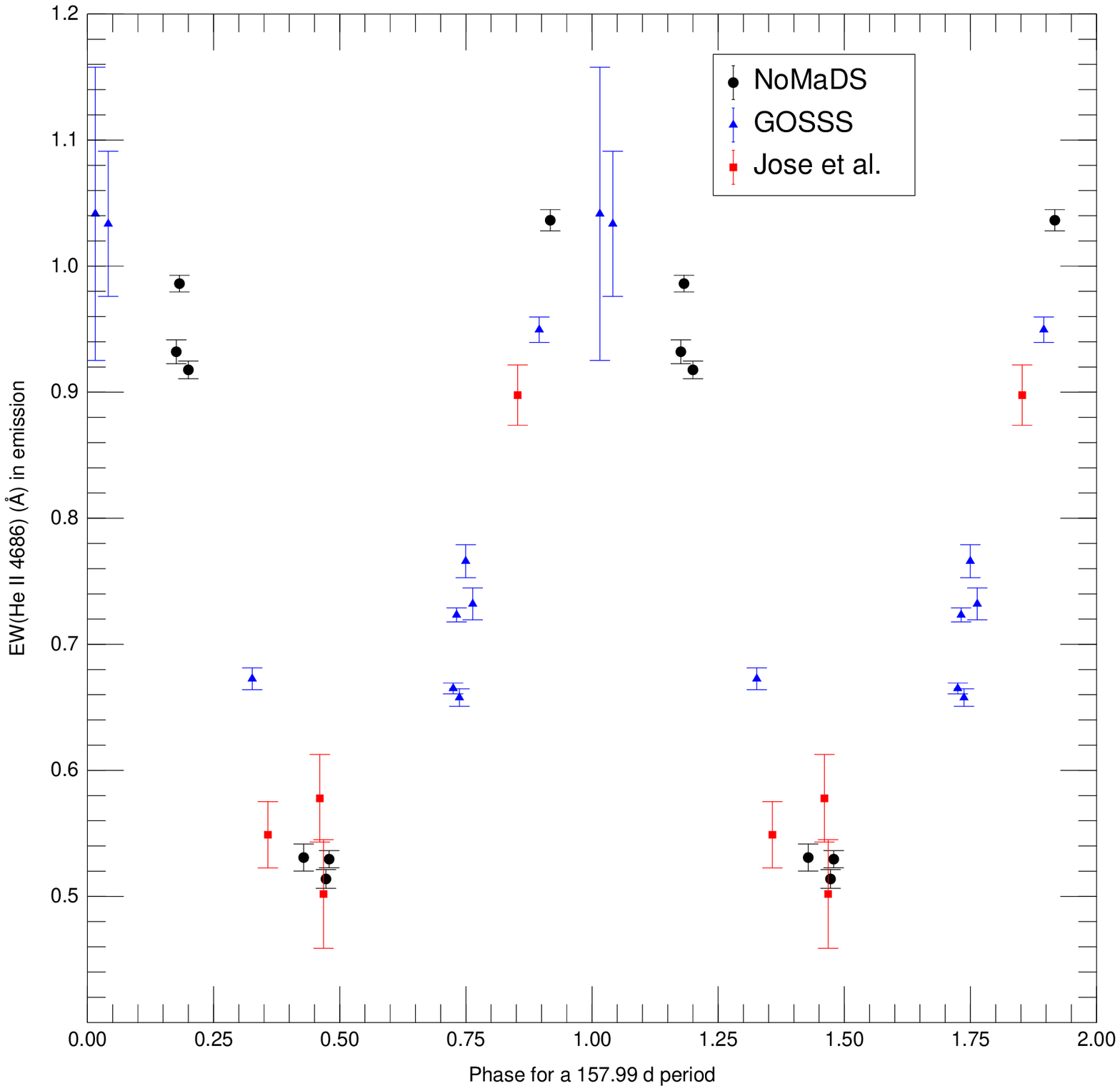}
\caption{\label{4686periodogram}{\em Left panel:}\ Periodogram of the He~{\sc ii} $\lambda 4686$ equivalent width measurements. {\em Right panel:}\ Equivalent widths of the He~{\sc ii} $\lambda 4686$ line of NGC 1624-2, phased according to Eq. (1). The ESPaDOnS observations of this star were acquired at phase $\sim 0.96$, while the Narval observation was acquired at phase $\sim 0.28$.}
\end{figure*}


\section{Spectroscopic variability and period}\label{period analysis}

The first indication of variability in the spectrum of NGC 1624-2 resulted from a comparison of the different GOSSS observations obtained
in 2008 and 2009 (see Table~\ref{spectroscopy}), reported by \citet{Walbetal10a} and \citet{Sotaetal11a}. Indeed,  \citet{Walbetal10a} hinted at the
possibility of variability on a time scale of days and suggested the need for further observations.

\begin{table}
\begin{center}
\caption{\label{he4886table}Equivalent width of He~{\sc ii} $\lambda 4686$) (in \AA) in emission as a function of Julian Date. The final column indicates the origin of the spectrum, as in Table~\ref{spectroscopy}.}
\begin{tabular}{cccc}
\hline
JD & EW & $\sigma_{\rm EW}$ & Type \\
(- 2\,450\,000) & (\AA) & (\AA)\\
\hline
3\,985.8 & 0.578 & 0.035 & J \\
3\,987.0 & 0.502 & 0.043 & J \\
4\,127.6 & 0.549 & 0.026 & J \\
4\,754.6 & 0.673 & 0.009 & G \\
5\,133.6 & 0.665 & 0.004 & G \\
5\,134.6 & 0.723 & 0.006 & G \\
5\,135.5 & 0.658 & 0.007 & G \\
5\,137.5 & 0.766 & 0.013 & G \\
5\,139.7 & 0.732 & 0.013 & G \\
5\,153.7 & 0.898 & 0.024 & J \\
5\,160.5 & 0.950 & 0.010 & G \\
5\,795.9 & 1.036 & 0.008 & N \\
5\,836.8 & 0.932 & 0.009 & N \\
5\,837.8 & 0.986 & 0.007 & N \\
5\,876.7 & 0.531 & 0.011 & N \\
5\,883.7 & 0.514 & 0.007 & N \\
5\,884.7 & 0.530 & 0.007 & N \\
5\,969.5 & 1.041 & 0.116 & G \\
5\,973.5 & 1.034 & 0.058 & G \\
5\,998.6 & 0.918 & 0.007 & N \\
\hline\hline
\end{tabular}
\end{center}
\end{table}

In order to determine the period of NGC 1624-2, we analyzed different optical lines and line ratios. We settled on four
lines: H$\alpha$, He\,{\sc i}~$\lambda$5876, H$\beta$, and He~{\sc ii}~$\lambda$4686; and one line ratio, 
He~{\sc ii}~$\lambda$4542/He\,{\sc i}~$\lambda$4471. Of these six lines, three remain in emission in all of our observations (H$\alpha$, 
H$\beta$, and He\,{\sc ii}~$\lambda$4686), two remain in absorption (He\,{\sc ii}~$\lambda$4542 and He\,{\sc i}~$\lambda$4471) and the last
one (He\,{\sc i}~$\lambda$5876) changes from emission to a complex absorption/emission profile. The variation of the He~{\sc i} $\lambda 5876$ line is
illustrated in Fig.~{\ref{5876}}. 

We measured the emission equivalent width (EW; i.e. the integral of $\delta\lambda\times$(flux-continuum)/continuum)) of
all the lines at every epoch and we analyzed the variations of the first four and the line ratio of the last two by means of Scargle
periodograms as described by \citet{1986ApJ...302..757H}. We examined periods in the range 60-400 days, using $10^5$ periods uniformly distributed in $\log {\rm (period)}$. In all five cases the periodograms show maxima between 156 and 159 days (Table~\ref{periods}). The periodogram for He\,{\sc ii}~$\lambda$4686 is shown in Fig.~\ref{4686periodogram} (left panel).  If we phase the EW measurements of this line with the adopted period (158 d, discussed below, and corresponding to the dominant peak of the peoriodogram), we obtained the variations shown in Fig.~\ref{4686periodogram} (right panel), where phase 0 has been set for 9 Feb 2012 at noon UT (i.e. HJD 2455967.0; these same measurements are summarized in Table~\ref{he4886table}). The behaviour
is the same for all the five quantities measured, with maxima (in either the EW or the line ratio) around phase 0.10-0.25 and minima around
phases 0.6-0.75. In addition, variations at the 0.01 magnitude level were detected in an $R$-band monitoring program of NGC 1624-2
with the 0.6 m telescope at the Esteve Duran Observatory that are consistent with the changes in EW(H$\alpha$) detected with spectroscopy. 

\begin{table}
\begin{center}
\caption{\label{periods}Periods derived from analysis of spectra line EW variations. The procedures used are described in the text. The adopted period is that corresponding to "Combined 1". The uncertainty quoted there is the dispersion of the five values, not the uncertainty of the
mean. Even though the five values appear to be compatible amongst themselves (their normalized dispersion with respect to the weighted mean is 1.6),
they are not fully independent measurements of the period, since the EWs and line ratios were not obtained at completely different times
for the different quantities (e.g. H$\beta$, He\,{\sc ii}~$\lambda$4686, and He\,{\sc ii}~$\lambda$4542~/~He\,{\sc i}~$\lambda$4471 were
obtained at exactly the same times).}
\begin{tabular}{lc}
\multicolumn{1}{c}{Quantity} & Period \\
\multicolumn{1}{c}{measured} & (d)    \\
\hline
H$\alpha$                                               & $156.57\pm0.53$ \\
He\,{\sc i}~$\lambda$5876                               & $158.38\pm1.70$ \\
H$\beta$                                                & $158.97\pm0.66$ \\
He\,{\sc ii}~$\lambda$4686                              & $158.52\pm0.52$ \\
He\,{\sc ii}~$\lambda$4542 / He\,{\sc i}~$\lambda$4471  & $158.31\pm0.65$ \\
Combined 1                                              & $157.99\pm0.94$ \\
Combined 2                                              & $156.87\pm2.95$ \\
\hline
\end{tabular}
\end{center}
\end{table}

\begin{figure*}
\centering
\includegraphics[width=5.9cm]{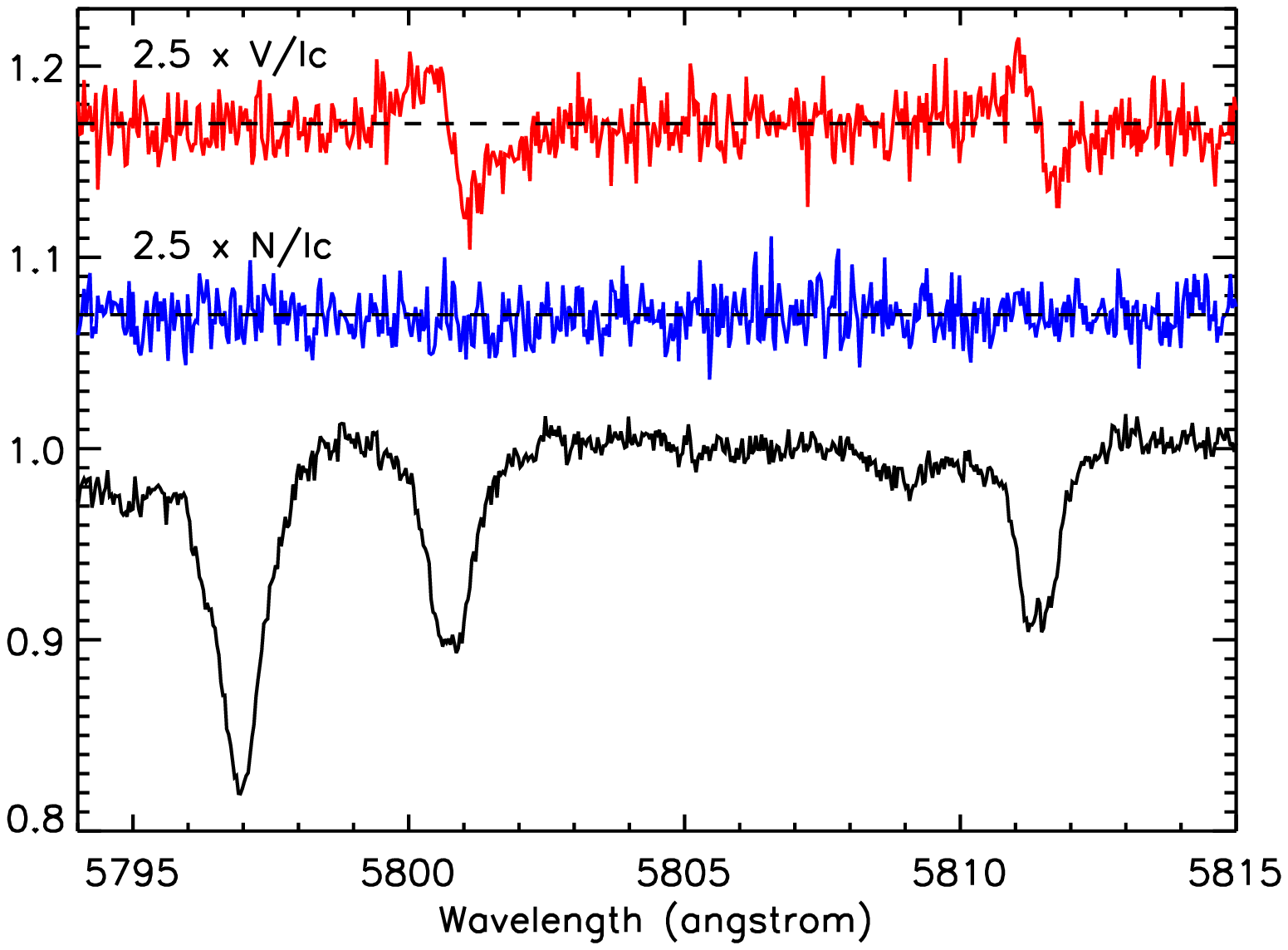}\includegraphics[width=5.9cm]{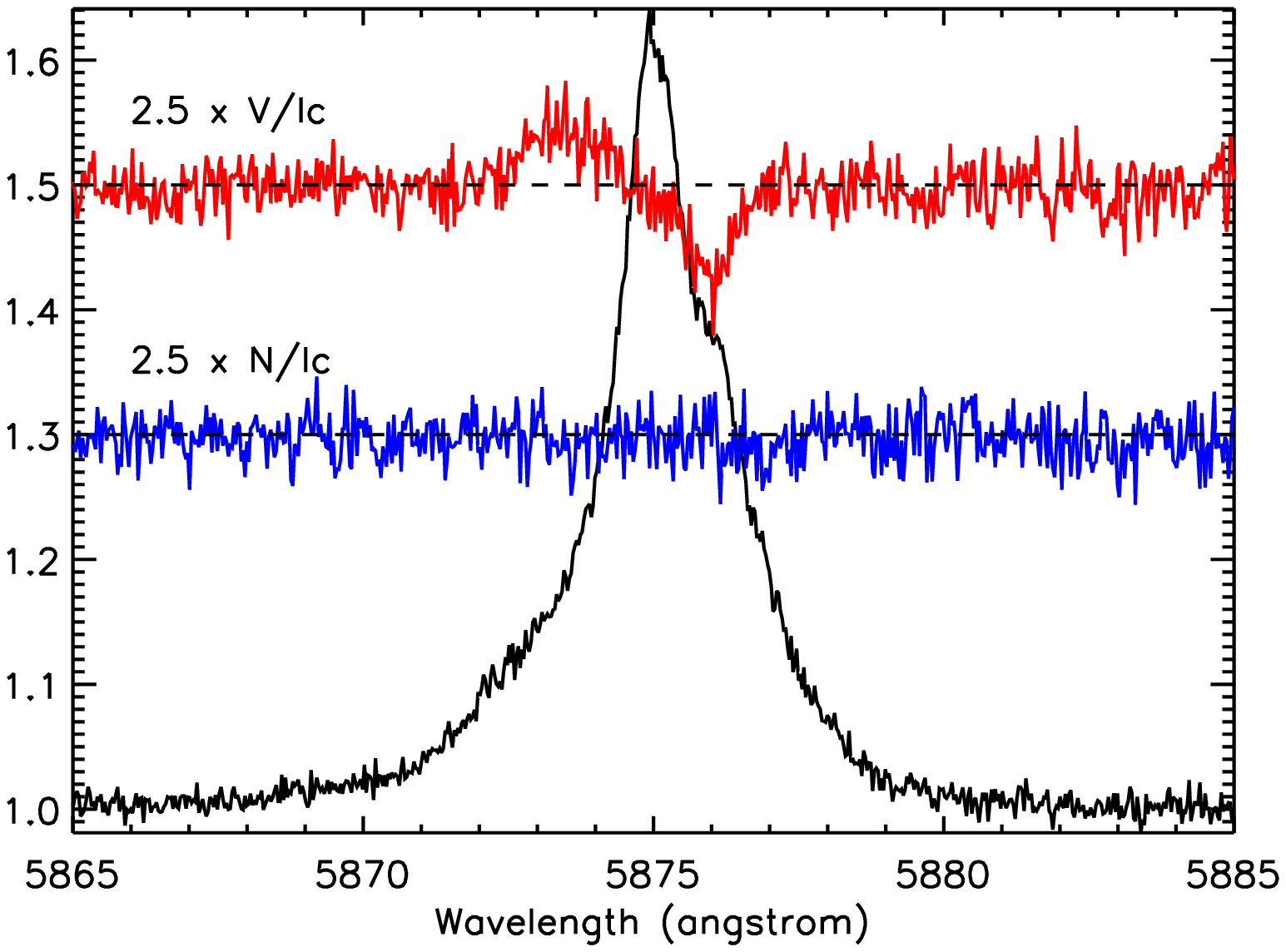}\includegraphics[width=5.9cm]{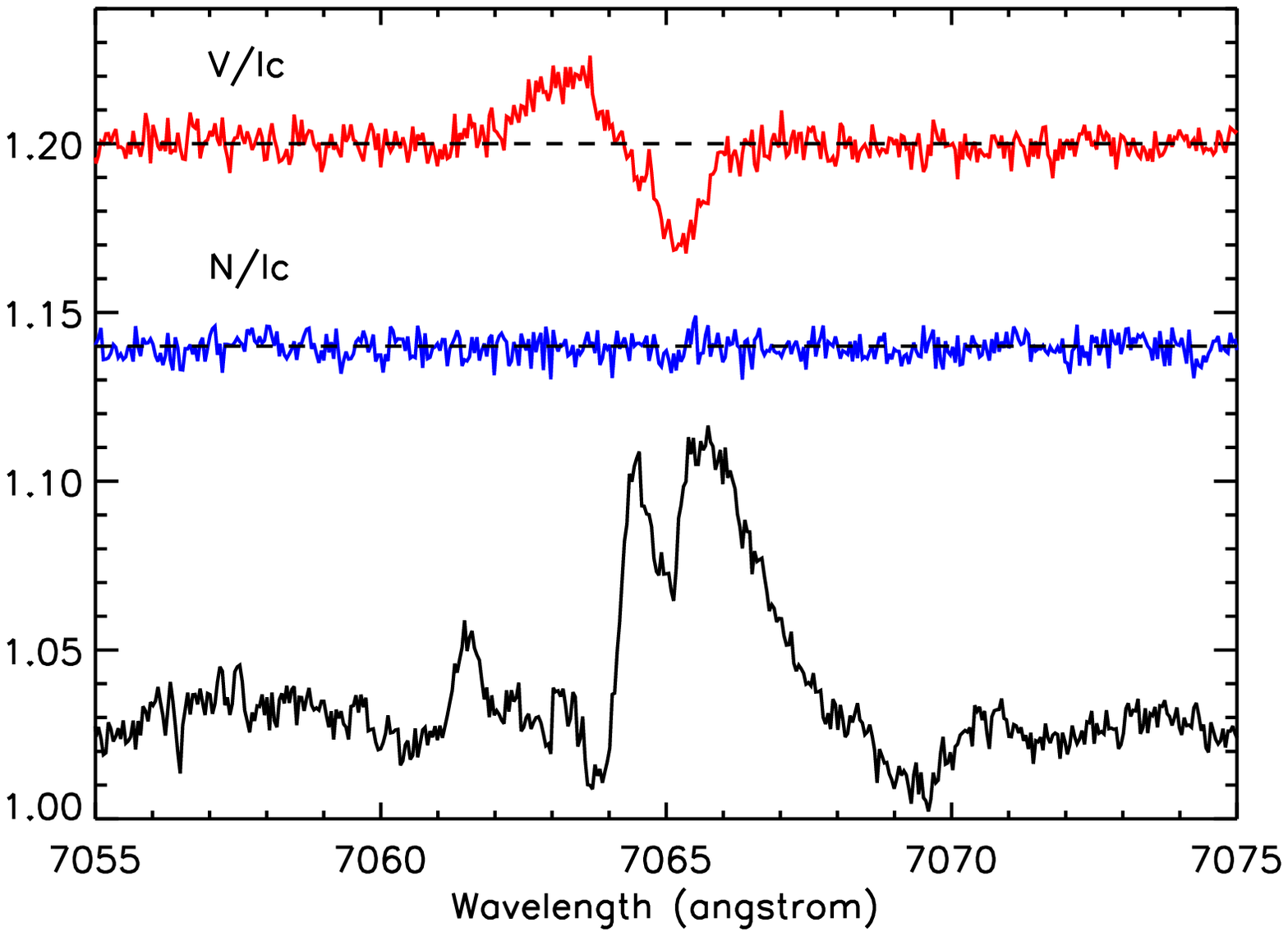}
\caption{\label{stokesvindlines}Stokes $I$ and $V$ profiles in the co-added ESPaDOnS spectrum of NGC 1624-2 (phase 0.96). {\em Left:} C~{\sc iv} $\lambda\lambda 5801, 5812$ lines, along with the Diffuse Interstellar Band (DIB) feature at 5797~\AA. {\em Middle:}\ He~{\sc i} $\lambda 5876$. {\em Right:}\ He~{\sc i}~$\lambda7066$. Note the clear Stokes $V$ signatures present in all of the stellar lines, and the lack of any signature in the DIB feature. Note $V$ and $N$ have been scaled (in the left and middle panels) and shifted for display purposes.}
\end{figure*}

\begin{table*}
\caption{Longitudinal field measurements obtained from individual lines in the ESPaDOnS and Narval spectra of NGC 1624-2, evaluated using Eq. (2). Uncertainties are formal errors computed by propagation of the spectral error bars. Listed are the heliocentric Julian date of the midpoint of the observation, {the instrument used,} the peak signal-to-noise ratio per 2.6~\kms\  velocity bin, and the evaluation of the longitudinal field from Stokes $V$. In no case is any detection obtained in the $N$ profiles. The first 5 rows provide the results for the individual ESPaDOnS spectra. The 6th row provides the statistical mean and standard deviation for the measurements of each spectral line. The 7th row provides the longitudinal field measured from the mean spectrum obtained from the co-addition of the 5 individual ESPaDOnS spectra. {The final row provides the results for the Narval spectrum.}\label{bzindlines}}
\begin{center}
\begin{tabular}{ccrrrrrrrrrrrrrrr}\hline\hline
HJD   & Inst       & SNR &            \multicolumn{8}{c}{$B_\ell\pm \sigma_B$ (kG)} \\
      -2455000   && (bin$^{-1}$)         &  C~{\sc iv}~$\lambda 5801$ &  C~{\sc iv}~$\lambda 5811$ &  He~{\sc i}~$\lambda 4712$ &  He~{\sc i}~$\lambda 4921$&  He~{\sc i}~$\lambda 5015$&  He~{\sc ii}~$\lambda 5411$&  He~{\sc i}~$\lambda 7281$&  O~{\sc iii}~$\lambda 5591$ \\
                             \hline
958.715 & E &    140&  $   6.62 \pm  1.4    $    &  $  8.27 \pm   1.1  $&  $  5.39  \pm  1.0  $ &  $ 7.46  \pm  1.1 $&  $   7.10  \pm   1.3  $&  $   3.50 \pm  1.1   $  &  $ 4.71 \pm   0.7  $&  $  8.91  \pm  1.3     $ \\
959.716 & E &    135 &  $   8.94 \pm   1.4    $ &  $  2.47 \pm   1.0  $&  $  3.21  \pm  1.0 $ &  $  5.49 \pm   1.2 $&  $   5.19  \pm   1.2   $&  $    1.89 \pm   1.3    $&  $ 4.34  \pm  0.8   $&  $ 8.51  \pm  1.5     $\\
960.713 &  E &   86& $   4.29 \pm   1.7    $   &  $  5.58 \pm   1.3  $&  $  3.80  \pm  1.6 $ &  $  9.70 \pm   1.5$&  $    9.06\pm     1.5$&  $    1.68 \pm   1.8 $   &  $ 6.56 \pm   0.9$  &  $  7.28  \pm  1.6   $\\
961.713 &  E &    137 &$   5.65 \pm   1.4    $ &  $  3.52 \pm   1.0  $&  $  2.95  \pm  1.0 $ &  $ 3.15  \pm  1.1 $&  $   8.08  \pm   1.1   $&  $    5.34 \pm   1.2    $&  $ 2.76 \pm   0.8 $ &  $  12.56 \pm   1.4   $\\
966.720 &  E &   116& $   8.47 \pm   1.6    $  &  $  4.06 \pm   1.1  $&  $  8.92  \pm  1.3 $ &  $  7.84 \pm   1.4$&  $    3.96 \pm    1.3 $&  $    5.85 \pm   1.4  $  &  $ 6.67 \pm   0.9 $ &  $  1.63  \pm  1.5    $  \\        
\hline\noalign{\smallskip}
$\overline{\langle B_{\rm z}\rangle}\pm\sigma$ & E &  - &   $   6.74 \pm   1.9    $ &  $4.77  \pm   2.3   $&  $  4.85\pm 2.5   $ &  $  6.73\pm 2.5 $&  $  6.68\pm    2.1  $&  $  3.65\pm  1.9   $  &  $  5.00\pm   1.6 $ &  $   7.78\pm   4.0   $\\
Co-added   & E &300&             $   6.59 \pm   0.8     $ &  $  4.75 \pm   0.6   $&  $ 4.66   \pm 0.7   $ &  $  6.30   \pm 0.6 $&  $    6.89 \pm    0.7  $&  $    3.73 \pm   0.6   $  &  $ 4.73 \pm   0.4 $ &  $  8.98 \pm   1.0   $\\
\hline\noalign{\smallskip}
1011.332 & N &   92   & $6.50\pm 1.6$ & $7.66\pm 1.4$ & $2.67\pm 1.4$ & $5.78\pm 1.0$ & $4.80\pm 1.7$ & $5.51\pm 1.6$ & $3.67\pm 1.0$ & $0.80\pm 0.6$ \\
\hline\end{tabular}
\end{center}
\end{table*}

To obtain a "best'' period we tried two different approaches. First, we combined the five measured periods weighted with their uncertainties
to obtain the sixth entry in Table~\ref{periods} ("Combined 1")\footnote{The uncertainty quoted there is the dispersion of the five values, not the uncertainty of the
mean. Even though the five values appear to be compatible among them (their normalized dispersion with respect to the weighted mean is 1.6),
they are not fully independent measurements of the period, since the EWs and line ratios were not obtained at completely different times
for the different quantities (e.g. the H$\beta$, He\,{\sc ii}~$\lambda$4686, and He\,{\sc ii}~$\lambda$4542~/~He\,{\sc i}~$\lambda$4471 were
obtained at exactly the same times).}. Secondly, before performing the period analysis we combined the various datasets by subtracting the mean value from each of the five quantities and dividing them by their respective dispersions. Then we calculated the periodogram from the combined dataset, yielding the last entry in Table~\ref{periods} ("Combined 2"). The uncertainty
there is significantly worse than for the other five, which we attribute to the fact that the phased profile variations for the five measured quantities 
are not exactly the same. 

We also checked that longer periods (e.g. twice the inferred period, i.e. $158\times 2=316$~d, corresponding to a double-wave variation) were incompatible with the data (in that they generated no coherent variation). No other acceptable phasing is found at this period or any other period.

Therefore, our preferred period is the sixth entry in Table~\ref{periods}, 157.99$\pm$0.94 days, and we adopt the ephemeris of the spectroscopic variations as follows:

\begin{equation}
{\rm JD}=(2455967.0\pm 10) +(157.99\pm 0.94)\cdot E.
\end{equation}

According to this ephemeris, the ESPaDOnS spectropolarimetric observations were acquired at phases 0.95-1.0, while the Narval data were obtained at phase 0.28.

If we assume that the adopted variability period corresponds to the period of rotation of the star \citep[as has been confirmed through the oblique rotator model for other Of?p stars: ][]{2011MNRAS.416.3160W,2012MNRAS.419.2459W}, we tentatively conclude that NGC 1624-2 is a very slowly rotating star, with a rotational period nearly 0.5 years in duration, and a single-wave variation of the EWs of its emission and absorption lines. Such slow rotation would be consistent with the negligible rotational broadening of spectral lines reported in Sect.~\ref{physicalproperties} and Sect.~\ref{magnetic}.


\section{Magnetic field}
\label{magnetic}

\begin{figure*}
\centering
\includegraphics[width=8.cm]{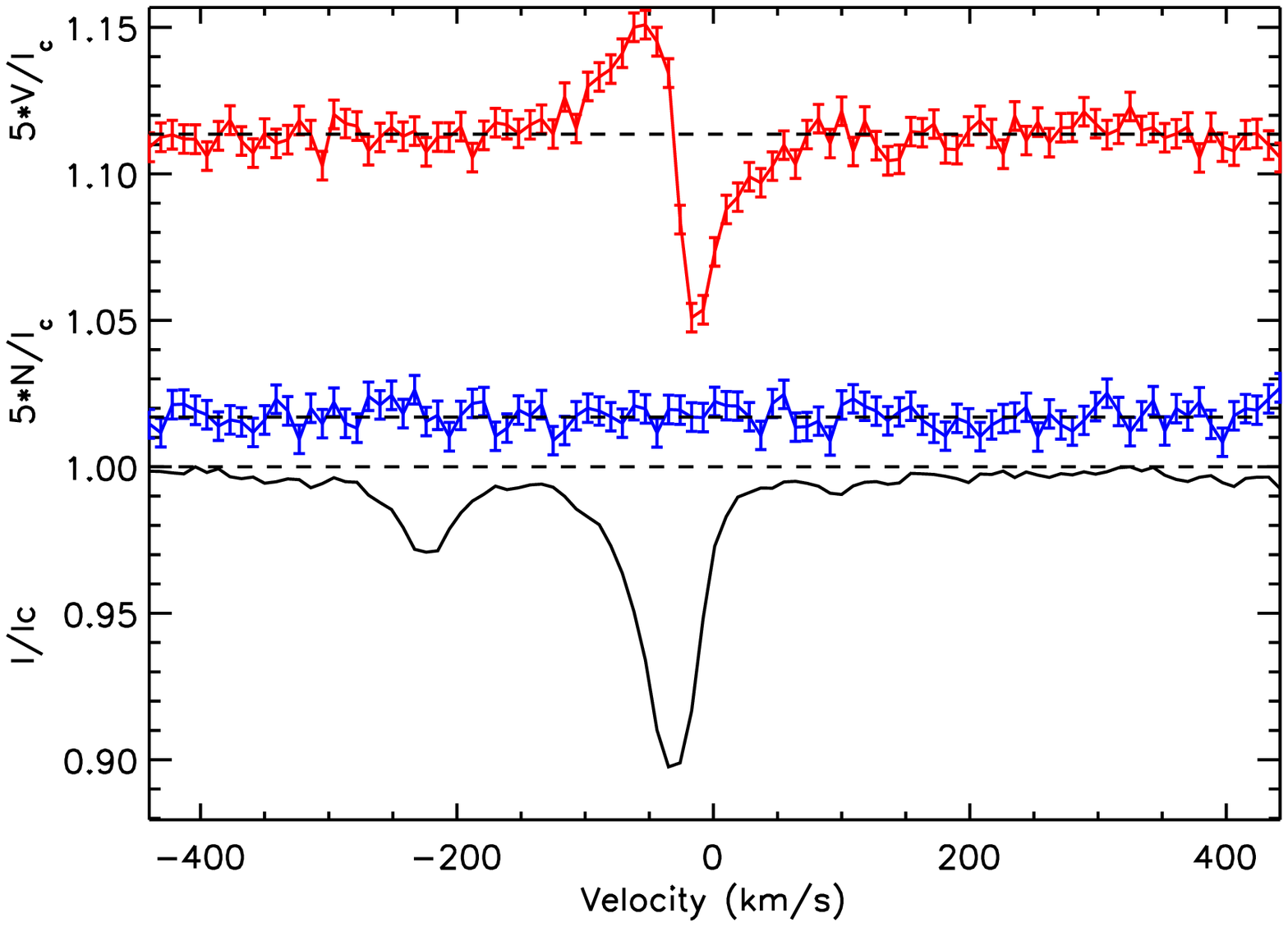}\includegraphics[width=8.cm]{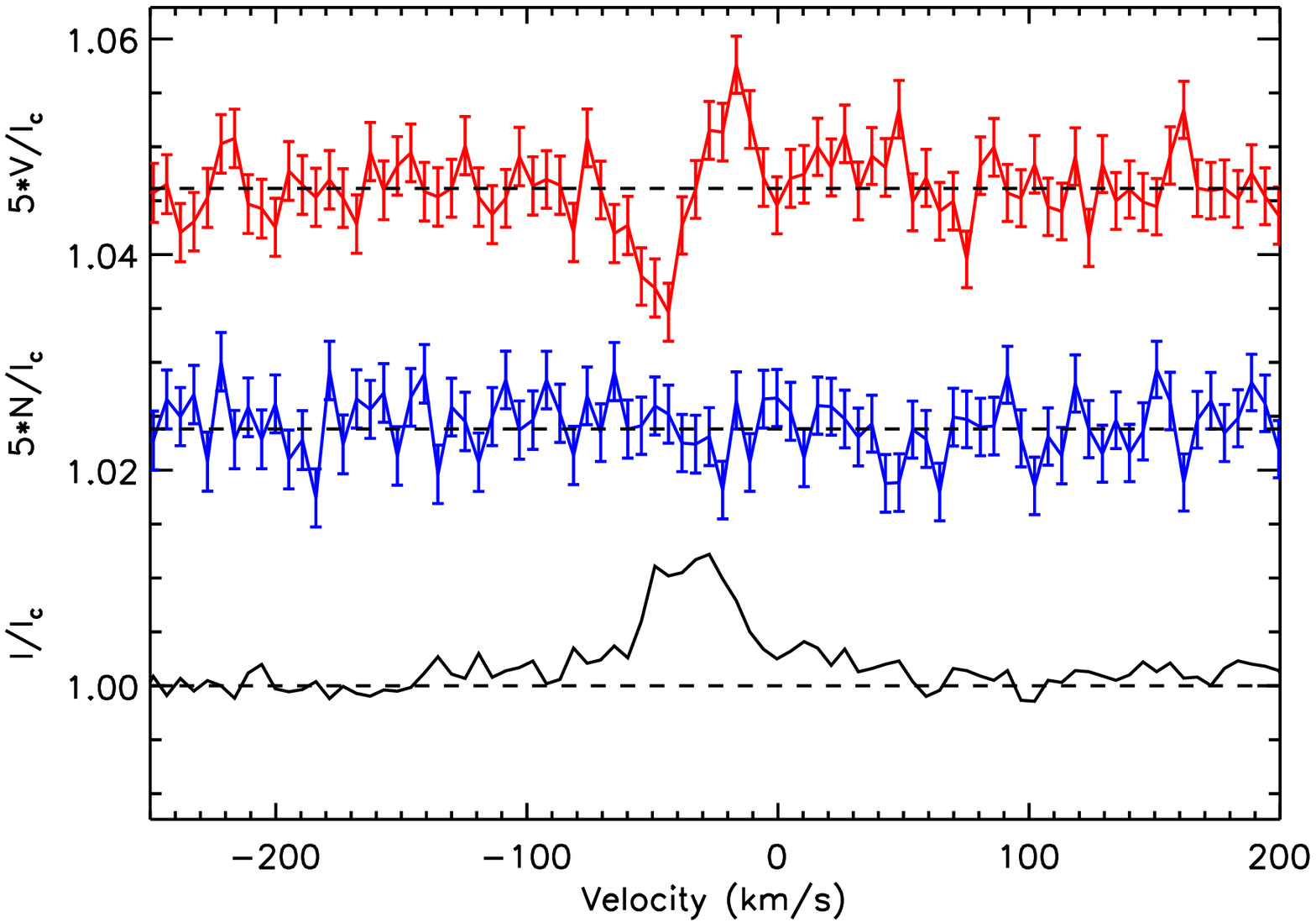}
\caption{\label{LSD}LSD profiles of NGC 1624-2. {\em Left:}\ LSD profiles from the first mask corresponding to absorption lines. The weak absorption feature at $\sim -200$~\kms is due to the DIB feature at 5797~\AA. {\em Right:}\ LSD profiles of weak O~{\sc iii} emission lines. Note the inverted Stokes $V$ profile relative to the left panel. This phenomenon is discussed in Sect. 6.4.}
\end{figure*}

The individual reduced ESPaDOnS and Narval Stokes $V$ spectra exhibit clear Zeeman signatures in profiles of various lines. Given the relatively low SNR ($\sim 100-150$) of the individual observations, this implies the presence of a very strong magnetic field. 

Examination of the ESPaDOnS Stokes $I$ spectra shows no significant variability of absorption or emission lines during the period of observation with that instrument (Feb 1-9 2012, phases 0.95-1.0 according to Eq. (1)). This is consistent with the long ($158$~d) period of spectral variability inferred for NGC 1624-2 (Sect.~{\ref{period analysis}). Taking advantage of the stability of the spectrum, we co-added the 5 ESPaDOnS observations. Examples of Zeeman signatures present in the C~{\sc iv} $\lambda\lambda 5801, 5812$, He~{\sc i}~$\lambda$5876 and He~{\sc i}~$\lambda$7065 lines of this mean spectrum are illustrated in Fig.~\ref{stokesvindlines}. According to Eq. (1), the flux-weighted phase corresponding to the co-added ESPaDOnS spectrum is 0.96.

\subsection{Longitudinal magnetic field from individual lines}

We began by measuring the longitudinal magnetic field \bz\ of individual spectral lines in each of the polarized spectra. We selected lines in absorption that appeared to be the least contaminated by emission and which showed clear Stokes $V$ signatures in the co-added spectrum: C~{\sc iv}~$\lambda\lambda 5801, 5811$, He~{\sc i}~$\lambda\lambda 4712, 4921, 5015, 7281$, He~{\sc ii}~$\lambda5411$,  and O~{\sc iii}~$\lambda 5591$. We used the first-moment method of \citet{1979A&A....74....1R}. All lines in the ESPaDOnS and Narval spectra are observed to have significant shifts from their rest wavelengths, corresponding to a radial velocity of about $-32$~\kms. We integrated the $I/I_{\rm c}$ and $V/I_{\rm c}$ profiles about their centres-of-gravity $v_{\rm 0}$ in velocity $v$, in the manner implemented by \citet{1997MNRAS.291..658D} and corrected by \citet{2000MNRAS.313..851W}:

\begin{equation}
\bz=-2.14\times 10^{11}\ \frac{{\displaystyle \int (v-v_{\rm 0}) V(v)\ dv}}{\displaystyle {\lambda z c\ \int [1-I(v)]\ dv}}.
\end{equation}

In Eq. (2) $V(v)$ and $I(v)$ are the $V/I_{\rm c}$ and $I/I_{\rm c}$ profiles, respectively. The wavelength $\lambda$ is expressed in nm and the longitudinal field \bz\ is in gauss. The wavelength and Land\'e factor $z$ correspond to those of each individual spectral line. Atomic data were obtained from the Vienna Atomic Line Database (VALD) where available. When experimental Land\'e factors were unavailable, they were calculated assuming L-S coupling. We integrated approximately symmetrically about $v_{\rm 0}$, using the observed span of the $I$ and $V$ profiles to establish the integration bounds. The \bz\ measurements obtained from each of the lines in the individual spectra, and the co-added ESPaDOnS spectrum, are summarized in Table~\ref{bzindlines}. The relevant atomic data are contained in Table~\ref{linemasks}. 

The grand mean obtained by averaging over all lines in Table~\ref{bzindlines} in the co-added ESPaDOnS spectrum is $5.83$~kG, with a standard deviation of 1.7~kG. This is consistent with the longitudinal field obtained from averaging the measurements from all lines in the individual ESPaDOnS spectra ($5.78$~kG with a standard deviation of $1.4$~kG). The Narval spectrum yields a similar mean value of \bz: 4.67~kG with a standard deviation of $2.2$~kG.The longitudinal magnetic field of NGC 1624-2 is clearly remarkably strong.

Table~\ref{bzindlines} shows that most lines exhibit somewhat more scatter in their measurements than is predicted by the formal errors (i.e. the standard deviation in row 6 is frequently $\sim 2$ times larger than the typical formal error of individual measurements). Examination of the measurements in greater detail revealed that the scatter results primarily from measurement of the first moment of $V/I_{\rm c}$, i.e. the numerator of Eq. (2). To explore the origin of the scatter, we first inspected the Stokes $V$ profiles of a given line, which we found to be identical within the noise. We then investigated the sensitivity of the measurements to adjustments of the integration range. We finally explored different methods of establishing $v_{\rm 0}$ (e.g. using only the Stokes $V$ profile, using Stokes $V$ weighted by the Stokes $I$ line depth, fixing $v_{\rm 0}$ at a constant value). Ultimately, we concluded that the extra fluctuations are likely due to the difficulty of establishing the position of the centre-of-gravity $v_{\rm 0}$ in the relatively noisy profiles of the individual spectra. This is supported by the measurements from the higher SNR co-added spectra (i.e. row 7 of Table~\ref{bzindlines}), which display no relatively extreme values, i.e. these values are in quite good agreement with the mean of the measurements for each line (row 6). We therefore conclude that measurements of \bz\ from individual lines in the individual spectra are susceptible to relatively large fluctuations, while the averaged measurements (or measurements obtained from higher SNR co-added spectra) are more robust.

We also observe that some spectral lines yield significantly lower/higher values of the longitudinal field than the mean (e.g. He~{\sc ii} $\lambda 5411$ ($3.73\pm 0.6$~kG in the co-added ESPaDOnS spectrum), vs. O~{\sc iii} $\lambda 5591$ ($8.98\pm 1.0$~kG)). This may result from several effects. For example, most lines in the spectrum of NGC 1624-2 are variable. The origin of this variability is not understood in detail; it may be a result of changes in the formation environment of a particular spectral line, or varying contributions of the photosphere vs. the magnetosphere. If lines are affected differently by these (potentially various, potentially complex) contributions, this may modify the numerator of Eq. (2) (e.g. if the line forms in a region where the magnetic field may be intrinsically weaker, e.g. the circumstellar environment) or the denominator (e.g. if an absorption line is infilled by unpolarised emission). A thorough investigation of these effects will require good-quality observations spanning the rotational cycle of the star, coupled with a more sophisticated understanding of the line formation processes and the origin of the variability. This star is uniquely well-suited to such an investigation.

\begin{figure*}
\centering
\includegraphics[width=8.5cm]{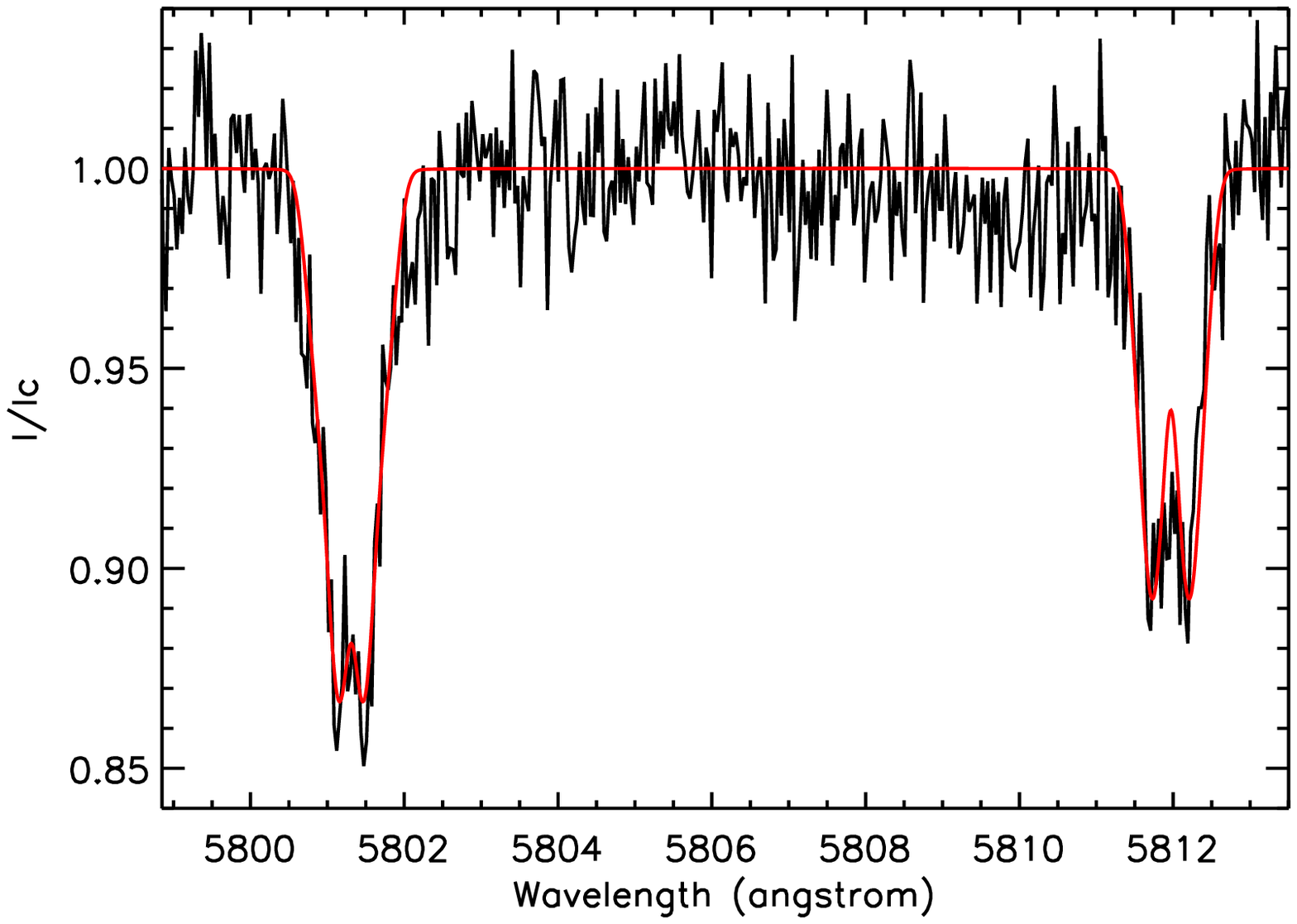}\includegraphics[width=8.5cm]{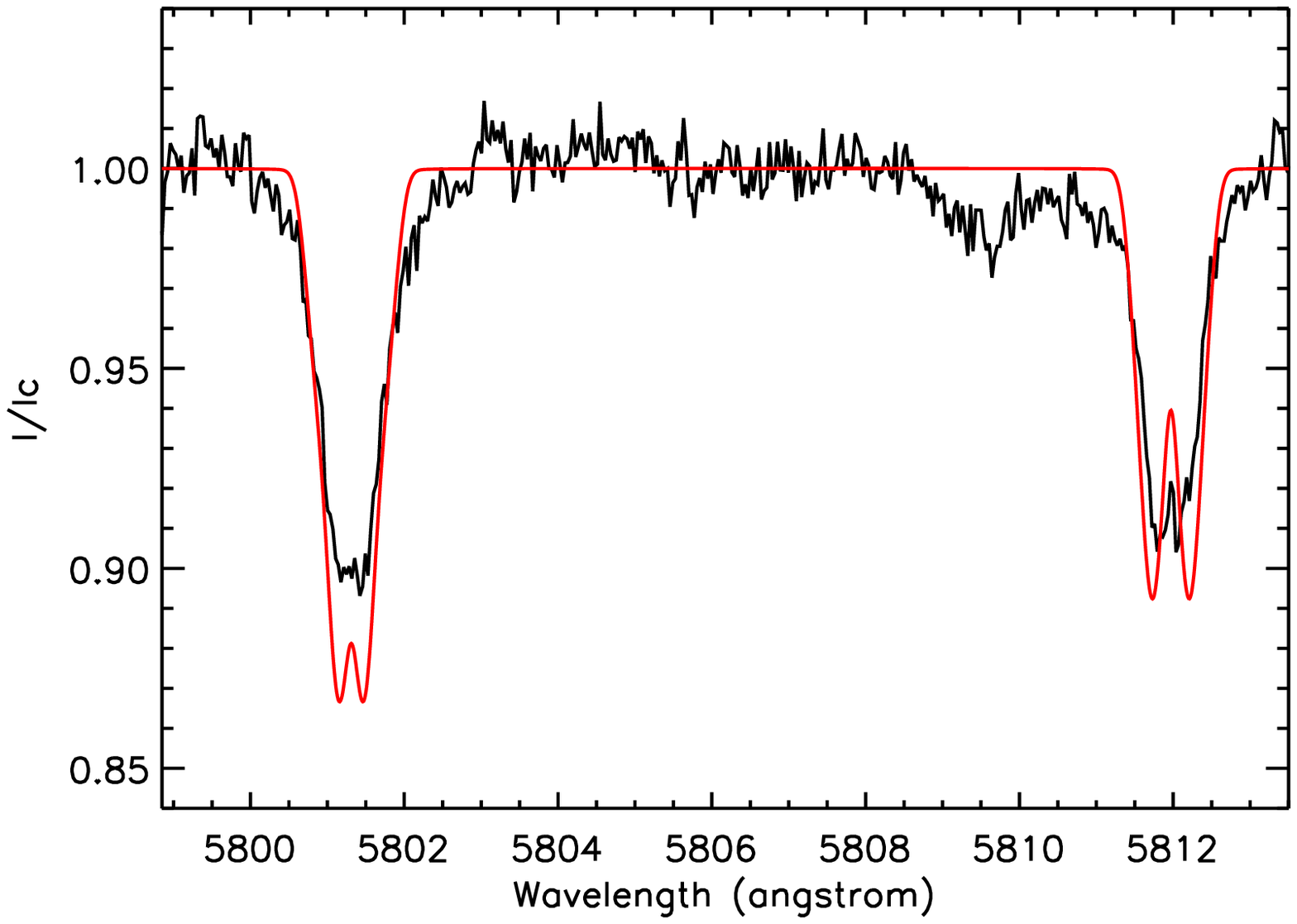}
\caption{\label{splitting}{\em Left -}\ Splitting of the Stokes $I$ profiles of the C~{\sc iv} $\lambda\lambda 5801, 5812$ lines in the Narval spectrum. Overplotted are model profiles for a $\sim 20$~kG dipole viewed at the positive magnetic pole. A solar abundance of C is assumed, and no additional broadening has been added. {\em Right -}\ The {\em same} theoretical profiles are compared to the co-added ESPaDOnS spectrum. Note the weaker lines and absence of resolved splitting in the ESPaDOnS observation. }
\end{figure*}

\subsection{Longitudinal magnetic field from LSD profiles}

We also analyzed the mean Stokes $I$ and $V$ profiles computed by means of Least-Squares Deconvolution \citep[LSD,][]{1997MNRAS.291..658D}. Donati's implementation of LSD was applied to the individual and co-added Stokes spectra. We took advantage of the unique characteristics of NGC 1624-2 - the detectability of Zeeman signatures in many individual lines - to develop a custom line mask for this star. The mask includes 8 spectral lines (not surprisingly, the same 8 spectral lines analyzed above), and is summarized in Table~\ref{linemasks}. 

Using the $\chi^2$ signal detection criteria described by \citet{1997MNRAS.291..658D}, we evaluated the significance of the signal in both LSD Stokes $V$ and in $N$. In no case is any signal detected in $N$, while signal in $V$ is detected definitely (false alarm probability ${\rm fap}<10^{-5}$) in all of our 5 ESPaDOnS observations, as well as in the Narval observation. From each set of LSD profiles we measured the mean longitudinal magnetic field in both $V$ and $N$ using Eq. (2), integrating from -125 to +90 \kms as determined from the radial velocity and extent of the Stokes $V$ and $I$ LSD profiles. We used the mean SNR-weighted Land\'e factor and wavelength, averaged over all lines in the mask. The longitudinal field measured from Stokes $V$ is detected significantly (i.e. $|z|=|B_\ell|/\sigma\ge 3$) in all but one of our observations, and in the mean ESPaDOnS spectrum at $10.7\sigma$ confidence ($\bz=5.35\pm 0.5$~kG). The longitudinal field in the Narval spectrum, $5.07\pm 1.2$~kG, is in formal agreement with the ESPaDOnS result obtained approximately 0.3 cycles earlier. In no case is the longitudinal field significantly detected in $N$. The results of the longitudinal field analysis are summarised in Table~\ref{LSDtable}. The LSD Stokes $I$, $V$ and $N$ profiles extracted from the co-added ESPaDOnS spectrum are illustrated in Fig.~{\ref{LSD}} (left frame). 

The longitudinal field inferred using LSD is formally consistent with that obtained from the individual line measurements from the co-added spectrum, using the standard deviation as the uncertainty. We note that the formal uncertainty associated with the LSD \bz\ measurement is substantially larger than the formal error we would compute from combining the co-added spectrum measurements (equal to 0.2~kG). This is a consequence of the inherent scaling applied to the LSD error bars in order to obtain a satisfactory agreement between the LSD polarized spectrum and the real observations \citep[see e.g.][]{2000MNRAS.313..851W, 2009MNRAS.398.1505S}. The LSD error bars therefore take into account line-to-line systematic errors such as those described in Sect. 6.1, and therefore probably provide a more realistic estimation of the uncertainties. As a consequence, we will adopt the LSD \bz\ measurements in the discussions that follow.

The strong longitudinal field inferred from the individual lines and LSD profiles fully supports our initial impression that the magnetic field of NGC 1624-2 is very strong. Assuming a limb darkening coefficient $u=0.3$ and a centred dipolar magnetic field, such a strong longitudinal field implies a (minimum) polar surface field strength of the dipole of nearly 20~kG according to Eq. (1) of \citet{1967ApJ...150..547P}. Such a field is nearly 8 times stronger than that of any other known magnetic O-type star.

\subsection{Zeeman splitting and magnetic field modulus}

As discussed in Sect.~{\ref{physicalproperties}}, metallic spectral lines of NGC 1624-2 are very sharp. Given the strong magnetic field inferred above, a significant component of this broadening may result from the magnetic field. In fact, while examination of the Stokes $I$ spectral lines in the ESPaDOnS spectra reveals no clear evidence of magnetic splitting, the C~{\sc iv} $\lambda\lambda$5801, 5812 lines in Narval spectra appear to show splitting in the line core (Fig.~\ref{splitting}, left panel).

To characterize the magnetic properties of NGC 1624-2 in more detail, to infer the surface field modulus (i.e. the disc-integrated modulus of the magnetic field, to which the Stokes $I$ Zeeman splitting is most straightforwardly related) and test the reality of the large longitudinal field discussed above, we attempted to reproduce the Stokes $I$ and $V$ profiles of C~{\sc iv}~$\lambda\lambda 5801, 5811$ using spectrum synthesis, taking into account the influence of the magnetic field. We used the Zeeman code \citep{1988ApJ...326..967L, 2001A&A...374..265W}, which computes profiles of spectral lines in LTE, taking into account Zeeman splitting and solving the equations of radiative transfer in all four Stokes parameters. We assumed a solar abundance ATLAS9 model atmosphere with $T_{\rm eff}=35$~kK and $\log g=4.0$. The assumption of LTE is probably rather poor in the case of such a hot star as NGC 1624-2, and likely questionable for inferring detailed abundances. Nevertheless, we expect that the splitting properties of spectral lines are not likely to be influenced in any significant way.


To begin we adopted a dipole magnetic field oriented such that our line-of-sight is aligned with the positive magnetic pole. Adjusting the local polar field strength to that predicted by Eq. (1) of \citet{1967ApJ...150..547P}, the Zeeman code reports a longitudinal magnetic field of 5.2~kG, in agreement with the measured field. The reported surface field modulus is 15~kG. As shown in Fig.~{\ref{splitting}} (left panel), {\em in the absence of any other line broadening}, this model does a reasonable job of reproducing the observed splitting and line width in the Narval spectrum. (We find in fact that while the best-fit field modulus to the $\lambda 5801$ line is 15~kG, for the $\lambda 5812$ line it is closer to 13~kG). This implies that the bulk of the rotational and turbulent broadening assumed in Sect.~\ref{physicalproperties} is likely magnetic. The Stokes $V$ profiles predicted by the model are in agreement with the observed circular polarization, although the Narval observations are sufficiently noisy that this agreement lacks any real meaning. From this comparison we conclude that the Stokes $I$ splitting observed in the Narval spectrum is consistent with the surface field expected from a nearly 20 kG magnetic dipole, and that in the presence of such a field other contributions to the broadening of the C~{\sc iv} lines (e.g. rotation, turbulence) are negligible (at least at the phase of the Narval observation).

Performing the same comparison with the co-added ESPaDOnS spectrum (Fig.~\ref{splitting}, right panel) we see that the C~{\sc iv} lines are weaker than in the Narval spectrum, and that no splitting is obvious. We find that this can be achieved in the models by reducing the field modulus and adding a small additional broadening (rotational, turbulent or magnetic) to reproduce the wings of the line profile. This corresponds to a total equivalent broadening of about 30~\kms. We find that the largest field moduli corresponding to dipoles that we are able to accommodate in this manner are about 12 kG for $\lambda 5812$, and 14~kG for $\lambda 5801$. Thus it appears that the field modulus inferred from the ESPaDOnS spectrum (at phase 0.96) is at least slightly weaker (upper limit of $13\pm 1$~kG) than that inferred from the Narval spectrum ($14\pm 1$~kG, at phase 0.28). 


\begin{table}
\caption{\label{linemasks}Atomic data of lines used for magnetic analysis, including for construction of line masks used for Least-Squares Deconvolution.}
\begin{center}
\begin{tabular}{ccrrrrrrrrrrrrrrr}
\hline\hline
 Wavelength & Species & $\chi_{\rm low}$ & Land\'e\\
(\AA) & & (eV) & factor \\
                             \hline
                             \multicolumn{4}{c}{Absorption lines}\\
4713.139 &   He~{\sc i}  &      20.964  &   1.250 \\       
 4713.156 &   He~{\sc i}  &      20.964  &   1.750 \\  
 4921.931 &   He~{\sc i}  &      21.218  &   1.000 \\  
 5015.678 &   He~{\sc i}  &      20.616  &   1.200 \\  
 5592.252 &   O~{\sc iii} &      33.858  &   1.000 \\  
 5801.312 &   C~{\sc iv}  &      37.549  &   1.167 \\  
 5811.968 &   C~{\sc iv}  &      37.549  &   1.333 \\  
 7281.349 &   He~{\sc i}  &      21.218  &   1.200 \\  
\hline 
                             \multicolumn{4}{c}{Emission lines}\\
7306.847  &  O~{\sc iii}    &   44.230  &   1.200   \\
 7307.117  &  O~{\sc iii}   &   44.243  &   1.750   \\
 7455.356  &  O~{\sc iii}   &   44.277  &   1.200   \\
 7515.987  &  O~{\sc iii}   &   44.277  &   2.250   \\
  8172.149  &  O~{\sc iii}  &   44.470  &   1.000   \\
\hline\hline\end{tabular}
\end{center}
\end{table}

\begin{table}
\caption{\label{LSDtable}Longitudinal field measurements of NGC 1624-2 obtained from LSD absorption line profiles. The final row provides the results for the mean spectrum obtained from coaddition of the 5 independent observations. In the "Inst" column, E=ESPaDOnS, N=Narval. In the "Det?" column, DD=Definite Detection.}
\begin{center}
\begin{tabular}{ccrrrrrrrrrrrrrrr}\hline\hline
    HJD      & Inst   & SNR      &    Det?  & LSD $B_\ell$ & LSD $N_\ell$ \\
-2455000  &    & pix$^{-1}$       &      &   (kG) & (kG)  \\
                             \hline
958.715 &E &140 &   DD& $5.95\pm 1.0$  & $-1.61\pm 0.9$ \\
959.716 &E &135 & DD &$6.33\pm 1.0$  & $-0.48\pm 1.0$ \\
960.713 & E &86  & DD & $6.01\pm 1.8$  & $-1.30\pm 1.8$ \\
961.713 & E &137 & DD & $4.76\pm 0.9$  & $-0.03\pm 1.0$ \\
966.720 & E &116 & DD &$3.34\pm 1.3$  & $+2.25\pm 1.3$ \\        
\hline\noalign{\smallskip}
Co-added  & E & 300 & DD &$5.35\pm 0.5$ & $-0.33\pm 0.5$ \\
ESPaDOnS \\
\hline\noalign{\smallskip}
1011.332 &N &  92 & DD &$5.07\pm 1.2$ & $+1.37\pm 1.2$\\   
\hline\end{tabular}
\end{center}
\end{table}

\subsection{Inverted Stokes $V$ profiles of weak emission lines}

The Stokes $I$ and $V$ spectra of NGC 1624-2 are quite complex. For example, Fig.~\ref{stokesvindlines} shows 5 absorption/emission features. Of the three absorption features, one is unpolarised (the (interstellar) DIB), and two display Stokes $V$ profiles consistent with a positive longitudinal field (i.e. with positive circularly-polarised flux in the blue wing). The two emission lines (He~{\sc i} $\lambda 5876$ and He~{\sc i} $\lambda 7065$) show Stokes $V$ profiles similar to those of the absorption lines. However, because these lines are in emission (i.e. they have negative equivalent widths), the longitudinal field that would be inferred from these lines would in fact be {\em negative}\footnote{This complex behaviour underscored the value of high spectral resolving power for investigating magnetic fields of objects like NGC 1624-2.}. 

During our exploration of the Stokes $V$ spectrum, we noticed that a number of weak emission lines of multiply-ionised light elements, primarily O~{\sc iii}, also exhibited Zeeman signatures. However, these signatures appeared to be inverted relative to those observed in the absorption lines. We constructed a second LSD mask consisting only of those lines (see Table~\ref{linemasks}), and extracted the mean LSD profile. As shown in Fig.~\ref{LSD} (right panel), our initial suspicions were verified. These weak emission lines do exhibit inverted Stokes $V$ signatures relative to the absorption lines and other strong emission lines such as He~{\sc i} $\lambda 5876$.  

According to our CMFGEN models, these lines should form very close to the photosphere in normal O-type stars. In the spectrum of NGC 1624-2, the width and radial velocity of these lines are consistent with those derived from the C~{\sc iv} absorption lines, suggesting that this is probably true is the present case.

The reversed signatures in these lines can be naturally explained if the $V$ profiles are formed in emission. This is in contrast to the other emission lines - "wind lines" - which exhibit $V$ signatures with signs that are consistent with those of the absorption lines. We interpret this to indicate that these latter signatures {\em do not} form in emission (i.e. in the emitting region). Rather, they would form primarily in the photosphere in absorption lines (where the source function is decreasing with height). The emission contribution of the magnetosphere and wind - dominating the observed lines - would presumably form at relatively large distances from the star, where the dipole field has decreased significantly in strength, and would therefore contribute only weakly to the observed $V$ profiles.

Measuring the longitudinal field from the O~{\sc iii} LSD profiles yields $\bz=2.58\pm 0.7$~kG. The longitudinal field from these emission lines is therefore consistent in sign with that measured from the absorption lines, but more than a factor of 2 smaller. However, as discussed above the emission lines are not predicted to probe substantially different spatial regions from the absorption lines, at least as estimated from a spherical wind model. An intriguing possibility that potentially explains the weaker longitudinal field strength has these lines forming primarily in the low-velocity plasma confined in closed magnetic loops, around $(5.35/2.58)^{1/3}-1\simeq 0.3~R_*$ above the stellar surface. Such a scenario is qualitatively consistent with the expected region of formation, the measured characteristics of the Stokes $I$ profiles (width and radial velocity), and the weaker measured longitudinal field of consistent sign. On the other hand, given the uncertainties related to the detailed formation of lines in the photosphere/magnetosphere/wind discussed in Sect. 6.1, we reserve further speculation until the line formation is better understood.

\section{X-rays}

NGC 1624-2 was observed for 10\,ks with the ACIS-I CCD array on the \textit{Chandra} X-ray Observatory in 2006 (ObsID 7473, PI Garmire, phase 0.12 according to Eq. 1), in which NGC\,1624-2 was detected with 38 counts. The source is located on-axis, but very close to a gap between the I2 and I3 CCDs. According to the exposure map provided by the \textit{Chandra} Source Catalogue \citep{2010ApJS..189...37E}, the unfortunate positioning of the source on the detector resulted in a loss of 35-40\% in effective area.

We extracted the source spectrum with \textsc{specextract} in CIAO~4.4, using a circular source region with a radius of 5 pixels ($\approx 2.5$\,arcsec), which encircles more than 95\% of the energy at an off-axis angle of 0.09\,arcmin.
The background was extracted from a larger, source-free region nearby, straddling the same fraction of the two chips. Despite the poor data quality, it is nonetheless clear that the spectrum is very hard, with only one count below 1\,keV. This suggests either very hot plasma, or a highly absorbed, somewhat cooler plasma.

We modelled the ACIS-I CCD spectrum in \textsc{xspec} 12.7 \citep{1996ASPC..101...17A} with a single-temperature, solar abundance  \citep{2009ARA&A..47..481A} APEC plasma \citep[with \textsc{atomdb} 2.0.1,][]{2001ApJ...556L..91S}, 
and the cold ISM absorption model of \citet{2000ApJ...542..914W}.
Given the low number of counts, we binned the spectrum from 0.5 to 7.0 keV in such a way as to obtain at least one count per bin (see Fig.~\ref{xray1}). We therefore used the C-statistic to determine the best-fit model and place constraints on the model parameters. 
We used an ISM column density $N_\mathrm{H}^\mathrm{ISM}=0.48\times10^{22}$\,cm$^{-2}$, determined from $E(B-V)=0.802$ and $R_\mathrm{V}=3.74$. For such a low column density, the resulting best fit value for the plasma temperature is quite high ($kT = 2.3^{+0.9}_{-0.3}$\,keV), although the fit is rather poor.

We then allowed for the possibility of extra \textit{local} absorption of the X-rays (from the wind or from a magnetosphere\footnote{Note that the local absorbing material may have different absorbing properties than assumed here \citep[see for example][]{2010ApJ...719.1767L}. }), in addition to the ISM value of $N_\mathrm{H}^\mathrm{ISM}=0.48\times10^{22}$\,cm$^{-2}$. The resulting fit, shown in Fig.~\ref{xray2}, is marginally better. The temperature is lower -- $kT = 0.7^{+0.4}_{-0.2}$\,keV -- and the necessary extra column density would be surprisingly high $N_\mathrm{H}^\mathrm{local}=1.9^{+0.7}_{-0.5}  \times10^{22}$\,cm$^{-2}$ compared to typical O-type stars \citep[e.g.][]{2011ApJS..194....7N}.
Fig.~\ref{xray2} shows the C-statistic 1, 2 and 3 $\sigma$ contours for the plasma temperature and local absorbing column, to illustrate the possible range in parameters. Higher SNR data are therefore needed to obtain robust constraints on the plasma temperature and absorbing column. 

The absorbed fluxes for the two models described above range from 3-6$\times10^{-14}$\,erg\,s$^{-1}$\,cm$^{-2}$. Correcting for the ISM column density only and using a distance of $5.2\pm1.1$\,kpc, the X-ray luminosity is 1-4$\times10^{32}$\,erg\,s$^{-1}$. With a luminosity of $\log(L/L_\odot)=5.1$, the X-ray efficiency is therefore $\log(L_\mathrm{X}/L_\mathrm{bol})\sim -6.4$, {a factor of about 4 higher} than the canonical value for O stars of $\sim -7$ \citep{2011ApJS..194....7N}.

High X-ray luminosity is a common feature among magnetic O-type stars with confined winds.
The Of?p stars HD\,108, HD\,191612 and HD\,148937, as well as O7\,V star $\theta^1$\,Ori\,C, have $\log(L_\mathrm{X}/L_\mathrm{bol})$ from $-6.0$ to $-6.2$ \citep{1989ApJ...342.1091C,2004A&A...417..667N,2005ApJ...628..986G,2007MNRAS.375..145N,2008AJ....135.1946N}.
However, these stars show a wide range of plasma temperatures. For the magnetic Of?p stars the plasma is dominated by a cool component similar to normal O-type stars, even if some hotter plasma is present. On the other hand, the X-rays of the O-type star $\theta^1$\,Ori\,C are dominated by the hard component.
At first glance, NGC\,1624-2 seems to have hotter plasma than normal O-type stars. However, a full, detailed modelling of NGC1624-2 must await higher SNR X-ray observations.

\begin{figure}
\includegraphics[width=84mm]{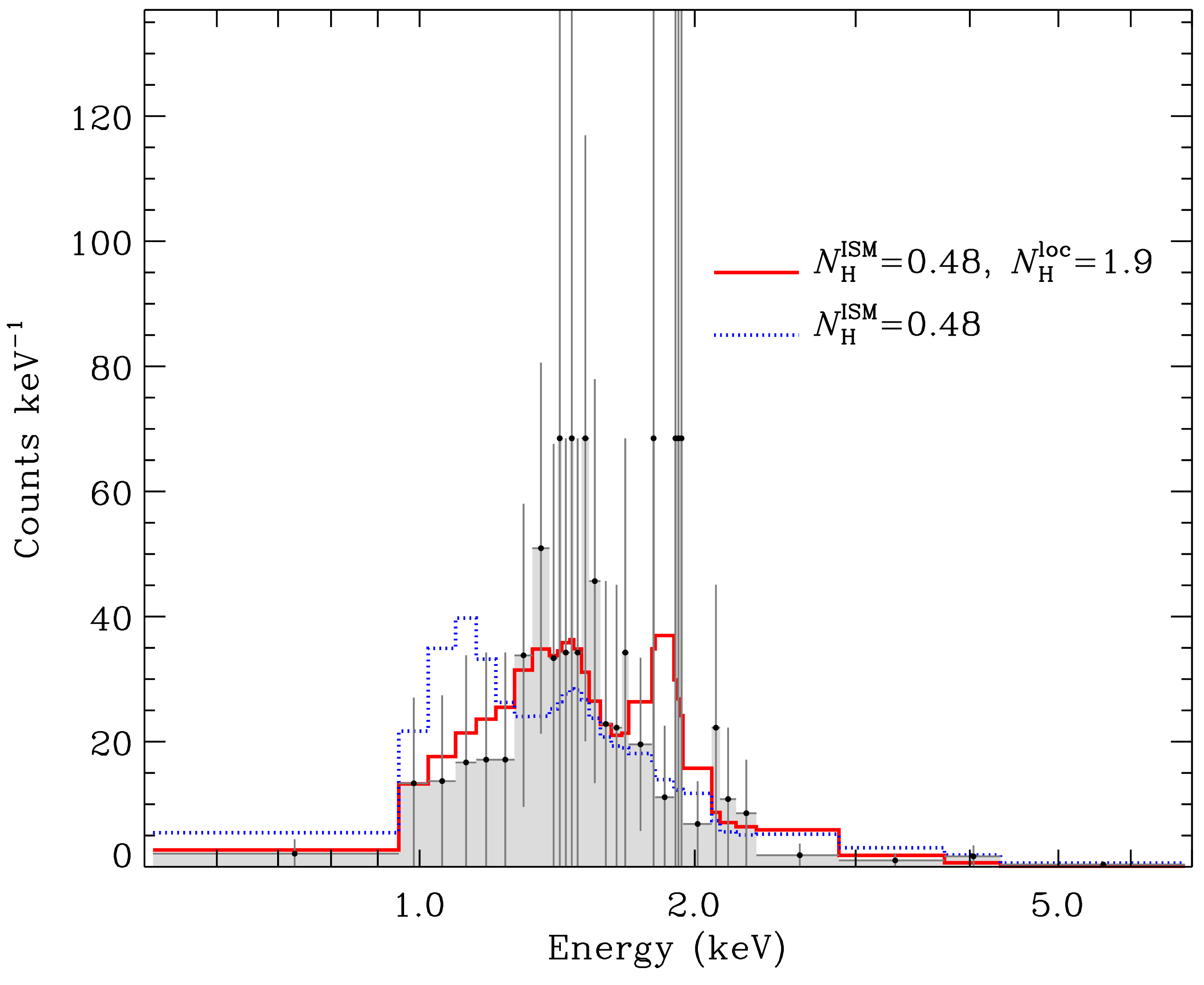}
\caption{\label{xray1}ACIS-I spectra of NGC\,1624-2, binned to obtain at least one count per bin (black dots, error bars and shaded area). The two models described in Table~\ref{xraytab} are shown in red (solid) and blue (dotted) lines. The C-statistics for the fits are 13.7 and 21.4, respectively.}
\end{figure}

\begin{figure}
\includegraphics[width=84mm]{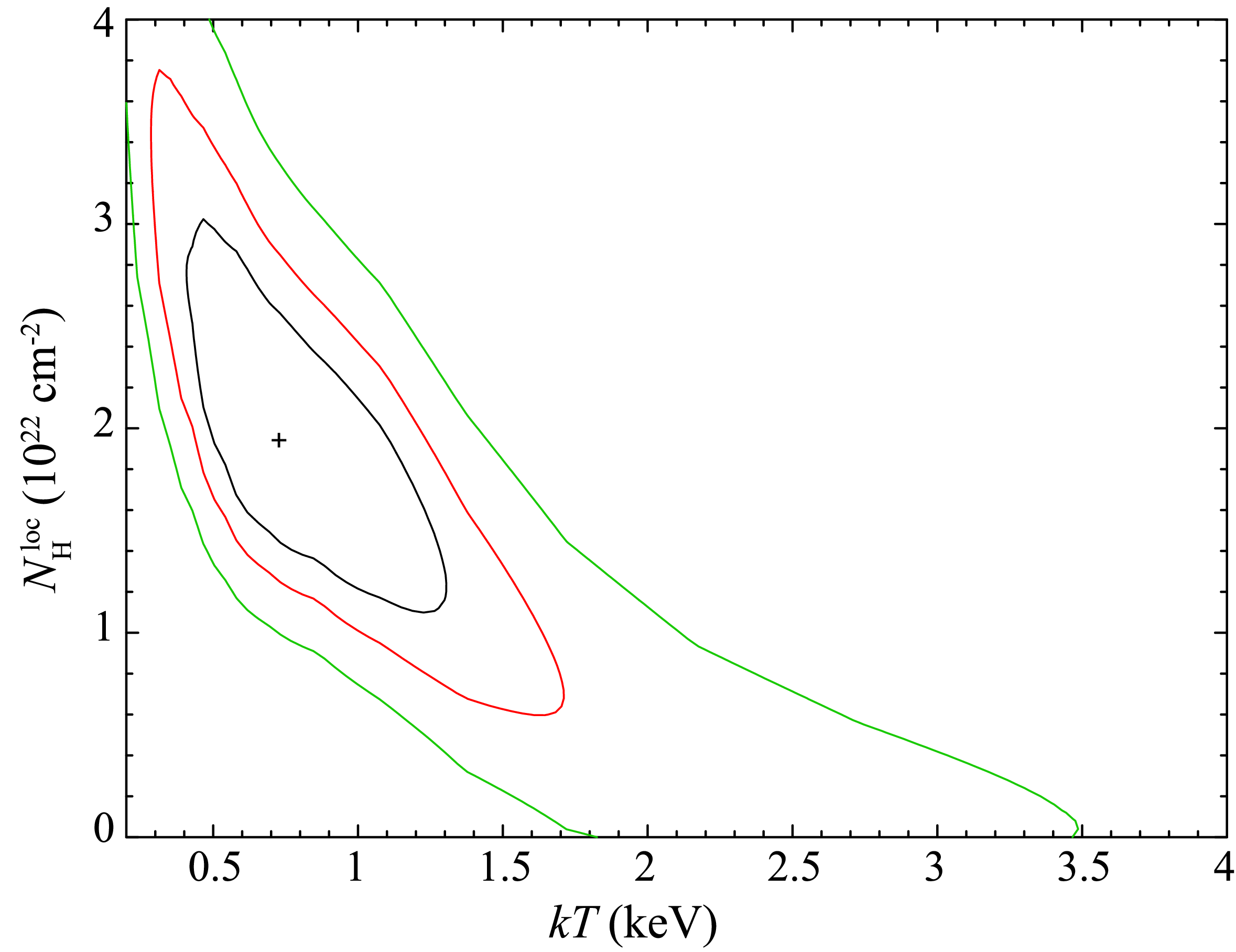}
\caption{\label{xray2}C-statistic 1, 2 and 3 $\sigma$ contours for the plasma temperature ($kT$) and local absorbing column ($N_\mathrm{H}^\mathrm{loc}$). The best fit is indicated by a cross.  }
\end{figure}

\begin{table}
\begin{center}
\caption{\label{xraytab} Single-temperature APEC plasma models, absorbed by a cool ISM model.}
\begin{tabular}{c c c c c} \hline
$N_\mathrm{H}^\mathrm{ISM}$ & $N_\mathrm{H}^\mathrm{local}$ & $kT$ & Abs. flux\,$^a$ & $L_\mathrm{X}$\,$^b$ \\
($10^{22}$\,cm$^{-2}$) & ($10^{22}$\,cm$^{-2}$) & (keV) & ($10^{-14}$\,erg\,s$^{-1}\,$cm$^{-2}$) & ($10^{32}$\,erg\,s$^{-1}$)\\
\hline
0.48 & 0 & $2.3^{+0.9}_{-0.3}$ & $5.0^{+2.0}_{-0.8}$ & $2.4^{+2.0}_{-1.4}$ \\
0.48 & $1.9^{+0.7}_{-0.5}$ & $0.7^{+0.4}_{-0.2}$ & $4.3^{+1.9}_{-1.2}$ & $1.8^{+1.6}_{-1.3}$ \\
\hline
\end{tabular}
\end{center}
$^a$ In the 0.5 - 7.0\,keV band.\\
$^b$ The absorbed flux was corrected for the $N_\mathrm{H}^\mathrm{ISM}$ only. 
\end{table}

\section{The giant magnetosphere of NGC 1624-2}
\label{magnetosphere}

Magnetospheres of O and early B stars form through the channelling and confinement of an outflowing wind by the star's magnetic field. 
This magnetic control breaks the symmetry of radiatively-driven winds, and therefore will influence observable wind diagnostics, such as Balmer line emission \citep{1978ApJ...224L...5L} and wind resonance lines in the UV \citep{1990ApJ...365..665S}. Furthermore, material forced to flow along the field lines will collide near the tops of closed loops, producing a shock-heated volume of plasma that will eventually cool, radiating X-rays \citep{1997ApJ...485L..29B,1997A&A...323..121B}. Given the large mass-loss rate expected for a 35\,kK O-type star, and the observed strong magnetic field, it is highly probably that such a structure exists around NGC 1624-2.

As presented by \citet{2002ApJ...576..413U}, the global competition between the magnetic field and stellar wind can be characterized by the so-called wind magnetic confinement parameter $\eta_\star \equiv B^2_\mathrm{eq}R^2_\star / \dot{M}v_\infty$, 
which depends on the star's equatorial surface field strength ($B_\mathrm{eq}$), stellar radius ($R_\star$), and wind momentum ($\dot{M}v_\infty$). For a dipolar field, one can identify an Alfv\'en radius $R_\mathrm{A}\simeq\eta_\star^{1/4}R_\star$, representing the extent of strong magnetic confinement of the wind. 
Above $R_\mathrm{A}$, the wind dominates and stretches open all field lines. But below $R_\mathrm{A}$, the wind material is trapped by closed field line loops, and in the absence of significant stellar rotation is pulled by gravity back onto the star on a dynamical (free-fall) time-scale.

To estimate the wind momentum of NGC 1624-2, we determined the theoretical wind terminal velocity given by:
\begin{equation}
	v_\infty = 2.6 v_\mathrm{esc}=2.6\sqrt{\frac{2GM(1-\Gamma_e)}{R_\star}},
\end{equation}

\noindent \citep{1995ApJ...455..269L} where $\Gamma_e \equiv \kappa_e L/4 \pi GMc$ is the standard Eddington parameter for electron scattering opacity  in a fully ionized wind $\kappa_e=0.34$\,cm$^2$\,g$^{-1}$. 
Using the adopted parameters given in Table~\ref{param_summary}, we obtain $v_\infty=2875$\,km\,s$^{-1}$.
We used the mass-loss rate recipe of \citet{2000A&A...362..295V,2001A&A...369..574V} and we obtain $\dot{M}=1.6\times10^{-7}$\,M$_\odot$\,yr$^{-1}$, roughly consistent with the other Of?p stars. Using a dipolar field strength of 20\,kG\footnote{While the maximum measured field modulus is $14\pm 1$\,kG, to remain consistent with $\eta_\star$ inferred for other magnetic O stars we employ for this purpose the equatorial field of a dipole of polar strength 20~kG inferred from the longitudinal field.}, this leads to a magnetic confinement $\eta_\star=1.5\times10^4$. The closed loop region of the magnetosphere therefore extends up to $R_\mathrm{Alf}=11\pm 4$\,R$_\star$ (assuming a conservative 40\% uncertainty in $B_{\rm d}$ (i.e. $\pm 5$~kG) and a factor of two uncertainty in wind momentum). Therefore, even taking into account the uncertainties, the magnetosphere of NGC\,1624-2 is intrinsically large, and substantially larger than that of any other magnetic O-type star (for all of which $R_\mathrm{A}<4\,R_\star)$. 

In the presence of significant stellar rotation, centrifugal forces can support any trapped material above a Kepler co-rotation radius $R_\mathrm{Kep}\equiv(GM/\omega^2)^{1/3}$. This requires that the magnetic confinement extend beyond this Kepler radius, in which case material can accumulate to form a centrifugal magnetosphere \citep[e.g.][]{2005ApJ...630L..81T}. In the case of NGC\,1624-2, the slow rotational period puts the Kepler radius much farther out (40\,R$_\star$) and no long-term accumulation of wind plasma is possible. However, the transient suspension of circumstellar material still results in a global over-density in the closed loops. For O-type stars with sufficient mass-loss rates, the resulting dynamical magnetosphere can therefore exhibit strong emission in Balmer recombination lines \citep{2011arXiv1111.1238P,2012MNRAS.tmpL.433S}.  For the huge volume of the magnetosphere of NGC 1624-2, we would expect particularly strong Balmer emission. In fact, as illustrated in Fig. 2 of \citet{Walbetal10a} and Fig.~\ref{halpha}, NGC\,1624 has - by a large margin - the strongest Balmer line emission of all known magnetic O-type stars. 

In the prototypical magnetic O-star $\theta^1$\,Ori\,C, the hard and luminous X-rays are seen to be rotationally modulated \citep{1997ApJ...478L..87G}. \citet{2005ApJ...628..986G} used 2D magnetohydrodynamical (MHD) simulations to show that the high temperature plasma can be well explained by shocks extending up to 1.5\,R$_\star$, comparable to its inferred Alv\'en radius of 2\,R$_\star$. Therefore, the trapped wind has only been accelerated to a fraction of its potential terminal velocity before reaching the top of the closed loops. 

In a larger magnetosphere such as NGC\,1624-2, the wind speed could in principle get higher before crossing the shock front and one could expect harder X-ray emission than in other magnetic O-type stars. Furthermore, if a larger fraction of the wind gets shock-heated, the overall emission level could be higher. 
However, the current X-ray observations do not provide sufficient constraints on the plasma temperature distribution, the total X-ray flux and the column density to be meaningfully modelled with sophisticated magnetosphere models.

\section{Discussion and conclusion}
\label{conclusion}

\begin{figure}
\includegraphics[width=85mm]{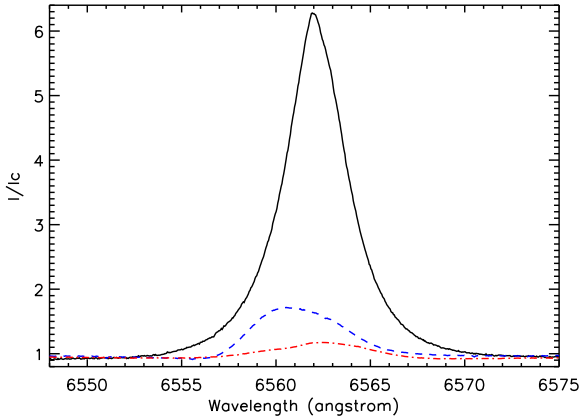}
\caption{\label{halpha}Peak H$\alpha$ emission of NGC 1624-2 (solid black line) compared to peak emission of HD 191612 (dashed blue line) and $\theta^1$~Ori C (dot-dashed red line). }
\end{figure}

We have discovered an extraordinarily strong magnetic field (maximum mean longitudinal magnetic field $\bz=5.35\pm 0.5$~kG, corresponding to a dipole of surface polar strength $\sim 20$~kG) in the Of?cp star NGC 1624-2 that distinguishes it qualitatively from other known magnetic O-type stars. Of particular interest is the presence of clear Zeeman signatures in individual spectral lines, and the apparent detection of resolved Zeeman splitting (corresponding to a maximum mean magnetic field modulus of $\langle B\rangle =14\pm 1$~kG). The detectability of Zeeman splitting should in principle allow a much stronger constraint on the geometry and topology of the star's magnetic field as compared to any other magnetic O-type star, once additional observations are acquired. We also suggest that NGC 1624-2 may be a good target for future transverse Zeeman effect (Stokes $Q$ and $U$) observations. With its sharp, magnetically-split lines, the $QU$ signatures could potentially have amplitudes comparable to Stokes $V$.

Using an extensive spectroscopic data set, we performed a first determination of the physical properties of the star. We confirm that it is a main sequence object with a mass comparable to those of the other Of?p stars. While the models used to infer the stellar properties included non-LTE effects, they did not directly include effects of the magnetic field. Such effects may include desaturation of profiles of spectral lines (modifying line blanketing), and introduction of Lorentz forces, both of which may lead to modification of the hydrostatic structure of the atmosphere \citep[e.g.][]{2009IAUS..259..407S, 2009IAUS..259..405S}. While the intrinsic uncertainties associated with the derived quantities are probably sufficiently large to dominate the observable consequences of these phenomena, considering the remarkable strength of the magnetic field of NGC 1624-2 their potential importance should be explored in more detail.

Indeed, the spectrum of NGC 1624-2 exhibits a number of peculiarities that distinguish this star from other Of?p stars. First, we have observed unprecedented composite profiles of the C~{\sc iii} $\lambda 4650$ complex, with narrow and broad components. We hypothesize that the broad components are photospheric as in normal Of or Ofc stars \citep{Walbetal10a}, while the narrow components typical of Of?p spectra are magnetospheric. Consistent with this interpretation, we observe that only the narrow components appear to vary with phase. Moreover, the narrow C~{\sc iii} emission does not disappear at minimum, so unlike most other members of this class, NGC~1624-2 remains Of?p at both extreme phases. We have also found that the absorption lines of NGC 1624-2 are very narrow, and that the widths of C~{\sc iv} $\lambda\lambda 5801, 5811$ can be reproduced essentially by magnetic broadening. This is in stark contrast to other Of?p stars for which the profiles of these lines are clearly dominated by turbulent broadening at the level of several 10s of \kms. The origin of this difference is currently unknown.  

We have also used the strong variations of various spectral absorption and emission lines to infer a unique and unambiguous spectral variation period of $157.99\pm 0.94$~d. Based on the observed behaviour of all other known magnetic O-type stars, it is reasonable to assume this to be the stellar rotation period (an assumption that {will} be tested by acquisition of additional magnetic field measurements). This implies that NGC 1624-2 rotates very slowly, spinning once in approximately one-half year. Such a conclusion is in good agreement with the negligible $v\sin i$ inferred from modelling the magnetically-split line profiles. It has been established that all known magnetic O-type stars have diverse but relatively long periods of rotation, from about a week \citep[HD 148937; ][]{2008AJ....135.1946N} to perhaps more than 50 years \citep[HD 108; ][]{2001A&A...372..195N}. The most common mechanism invoked to produce such slow rotation is magnetic braking, i.e. the shedding of rotational angular momentum via the stellar wind and enhanced lever arm provided by the magnetic field. If we employ the braking model described by \citet{2009MNRAS.392.1022U} (Eq. 25 of that paper), we can {roughly} compute the braking timescale $\tau_{\rm spin}$ as a function of the magnetic field strength, stellar mass and radius, mass loss rate and terminal velocity. Using the physical parameters reported in Table~\ref{param_summary} and moment of inertia coefficient $k\sim0.1$ \citep{2004A&A...424..919C}, we obtain a spin-down time for NGC 1624-2 of 0.24\,Myr. As discussed in Sect. 1, the estimated maximum age of the cluster NGC 1624 is no greater than 4~Myr, i.e. no more than $\sim 17\tau_{\rm spin}$. Such an age is, however, easily sufficient for the star to have braked from an initial short rotation period to its current very long period. (For example, if we assume the star was initially rotating at critical ($P_{\rm crit}=1.1$~d), the time required to slow the rotation to 158~d would be just $\ln (158/1.1)=4.96\tau_{\rm spin}=1.2$~Myr.) {On the other hand, if the cluster age is significantly younger than 4 Myr this would place constraints on the initial rotational speed of the star (requiring a slower initial rotation) or a reconsideration of the origin of the current slow rotation. These conclusions are subject to the assumptions and limitations of the braking model (e.g. aligned magnetic and rotation axes computed in 2D) and the uncertainties of the input parameters ($B_{\rm d}$, $M_*$, $R_*$, $\dot M$, $v_\infty$ and $k$).}

{In our analysis of the optical spectrum of NGC 1624-2, we derived a upper limit on the photospheric N abundance: {([N/H]$\ltsim$0.3)}. We consider this upper limit to be uncertain due to our present inability to directly include the influence of the strong magnetic field on the line formation in our NLTE spectrum synthesis model. Accurate knowledge of the surface N abundance represents an important constraint on the interior rotation profile of a magnetic early-type star. As reported by \citet{2011A&A...525L..11M}, when magnetic braking occurs in a massive star characterized by internal differential rotation, a strong and rapid mixing occurs in layers near the surface. This results in enhanced mixing, resulting in enriched surface abundances of nitrogen relative to similar models with no magnetic braking. However, when solid-body rotation is imposed in the interior, the star is slowed so rapidly that surface enrichments are in fact smaller than in similar models with no magnetic braking. Therefore magnetic braking can enhance the surface N abundance or, in contrast, have little effect on it, depending on the internal rotation profile of the star. \citet{2012A&A...538A..29M} investigated the N surface enrichment of 6 known magnetic O-type stars, including 3 Of?p stars. Depending on the assumed initial rotation velocity of the Of?p stars, they found that that they display surface nitrogen abundances consistent with those of non-magnetic O stars (if their rotation was initially rapid, $v_{\rm rot}\sim 300$~\kms), or enhanced relative no non-magnetic O stars (if their rotation was initially modest, $v_{\rm rot}\sim$ a few times $10$~\kms). Assuming NGC 1624-2 has a main sequence age no greater than 4 Myr, comparison of our preliminary N abundance with the figures presented by \citet{2012A&A...538A..29M} indicates that (1) NGC 1624-2 has a N abundance slightly lower than non-magnetic O stars with similar positions on the HR diagram; (2) that the abundance is low relative to that expected for a star of this age and mass if it was rotating initially at high velocity; (3) but consistent with the expected abundance if the star was rotating initially at lower velocity. However, as discussed above these results are quite tentative, and require more detailed modelling and a more robust determination of the age of the star before firm conclusions can be drawn.}

As a consequence of its intense magnetic field, NGC 1624-2 is expected to host a magnetospheric volume substantially larger than any other magnetic O-type star. The inferred magnetic wind confinement parameter, $\eta_\star=1.5\times 10^4$, is 300 times larger than that of the Of?p star with the next-strongest field, HD 191612. This leads to a predicted Alfven radius of 11.4~$R_\star$ (versus 2.2~$R_*$ in the case of HD 191612). This much larger volume of confined plasma should result in much stronger magnetospheric emission \citep[e.g. according to the mechanisms discussed by][]{2012MNRAS.tmpL.433S}. Indeed, the H$\alpha$ emission of NGC 1624-2 is found to be substantially stronger than that observed in any other magnetic O-type star. For comparison, and as illustrated in Fig.~\ref{halpha}, the maximum EW of H$\alpha$ in the spectrum of HD 191612 is about 4~\AA\ \citep[e.g.][]{2011MNRAS.416.3160W}, while that of $\theta^1$~Ori C is about 2~\AA\ \citep[][]{2008A&A...487..323S}. The peak EW of the H$\alpha$ line of NGC 1624-2 is {26~\AA}, {\em 6.5 times greater than HD 191612 and 13 times greater than that of $\theta^1$~Ori C.} Modeling of this remarkable H$\alpha$ emission is an urgent priority, and first attempts are underway by Sundqvist, ud Doula et al. (priv. comm.).

It is expected that such strong magnetic wind confinement in the presence of such a powerful wind should lead to intense X-ray emitting shocks. Analysis of archival {\em Chandra} ACIS-I X-ray observations indicates a hard and luminous X-ray spectrum ($\log(L_\mathrm{X}/L_\mathrm{bol})\sim -6.4$), qualitatively consistent with theoretical expectations as well as the behaviour of $\theta^1$~Ori C. However, the current observations are more or less equally consistent with a relatively hard, weakly extinguished source, or a relatively soft, highly extinguished source. New higher-quality X-ray observations will be required to draw useful quantitative conclusions about the X-ray properties of NGC 1624-2.

Due to the paucity of magnetic data, the magnetic topology and geometry of NGC 1624-2 are at present only weakly constrained. Although the geometry (i.e. the inclination angle $i$ and the magnetic obliquity $\beta$) is normally inferred from the variation of the longitudinal magnetic field, observations of other magnetic O stars (e.g. $\theta^1$~Ori C, HD 191612, HD 57682) show that a clear relationship exists between the EW variations of optical emission lines diagnostic of the wind (e.g. H$\alpha$, He~{\sc i} $\lambda 5876$, He~{\sc ii} $\lambda 4686$) and the longitudinal field variation. In particular, the emission extrema of these stars all correspond to extrema of the (sinusoidally-varying) longitudinal field, with the emission maximum corresponding to the maximum unsigned longitudinal field. If we assume that the magnetic topology of NGC 1624-2 is roughly dipolar, we can expect with reasonable confidence that the ESPaDOnS observation, acquired at phase 0.96 near emission maximum, corresponds to the approximate maximum of the longitudinal field. The longitudinal field measured from the Narval observation, acquired at phase 0.28, is not very different from that of the ESPaDOnS measurement, suggesting that the variation of the longitudinal field is not very large, and moreover that it does not change sign as the star rotates. This implies that we view essentially only one magnetic hemisphere during the rotation of the star, and consequently $i+\beta<90\degr$. This conclusion is supported by the single-wave nature of the EW variations. When both magnetic hemispheres are visible during the stellar rotation \citep[i.e. $i+\beta\gg 90\degr$, as in the case of HD 57682; ][ and MNRAS, submitted]{2009MNRAS.400L..94G}, the EW variations of the emission lines exhibit a double-wave variation. In contrast, smaller values of the sum $i+\beta$ produce single-wave variations \citep[e.g. $\theta^1$~Ori C, HD 191612; ][]{2008A&A...487..323S,2007MNRAS.381..433H}. 

Clearly, NGC 1624-2 holds a special place amongst the known magnetic O-type stars. Its extreme magnetic field and puzzling spectral peculiarities have much to teach us, and demand immediate attention. Plans for observational follow-up - including optical spectroscopy, spectropolarimetry, and photometry; UV spectroscopy; and X-ray spectroscopy - are already underway, as are theoretical investigations of its H$\alpha$ and X-ray properties to better understand the magnetospheric geometry, structure and energetics.


\section*{Acknowledgments}
GAW and AFJM acknowledge support from the Natural Science and Engineering Research Council of Canada (NSERC). JMA acknowledges support from [a] the Spanish Ministerio de Ciencia e Innovaci\'on through grants AYA2010-15081 and AYA2010-17631; [b] the Junta de Andaluc\'{\i}a grant P08-TIC-4075; and [c] the George P. and Cynthia Woods Mitchell Institute for Fundamental Physics and Astronomy. YN acknowledges support from the Fonds National de la Recherche Scientifique (Belgium), the Communaut\'e Fran\c aise de Belgique, the PRODEX XMM and Integral contracts, and the ‘Action de Recherche Concert\'ee’ (CFWB-Acad\'emie Wallonie Europe). VP acknowledges support from the Fonds Qu\'ebecois de la Recherche sur la Nature et les Technologies. STScI is operated by AURA, Inc., under NASA contract NAS5-26555. CFHT and TBL observations were acquired thanks for generous allocations of observing time within the context of the MiMeS Large Programs.


\bibliography{ngc1624-2}

\end{document}